\documentclass[12pt]{article}
\usepackage{amssymb}
\usepackage{color}
\usepackage{epsfig,amssymb,amsfonts,amsmath,graphicx,dsfont,cite,xfrac}
\usepackage{authblk}
\usepackage{subcaption}
\usepackage{physics}
\definecolor{mygray}{gray}{0.5}
\usepackage[T1]{fontenc}
\usepackage{lmodern}

\usepackage{cite}
\usepackage[colorlinks=true,linkcolor=blue,citecolor=red]{hyperref}
\usepackage{soul,cancel}

\newcommand{\be}{\begin{equation}}
\newcommand{\ee}{\end{equation}}
\newcommand{\bea}{\begin{eqnarray}}
\newcommand{\eea}{\end{eqnarray}}

\title{A geometric formulation to measure global and genuine entanglement in three-qubit systems}

\author[${1}$]{Salvio Luna-Hern\'andez}
\author[${2}$]{Marco Enr\'iquez}
\author[${1}$]{Oscar Rosas-Ortiz}

\affil[${1}$]{\footnotesize Physics Department, Cinvestav, AP 14-740, 07000 M\'exico City, Mexico}
\affil[${2}$]{\footnotesize Tecnologico de Monterrey, School of Engineering and Sciences, 01389, Santa Fe, Mexico}

\date{}
\begin{document}

\maketitle

\begin{abstract}
We introduce a purely geometric formulation for two different measures addressed to quantify the entanglement between different parts of a tripartite qubit system. Our approach considers the entanglement--polytope defined by the smallest eigenvalues of the reduced density matrices of the qubit-components. The measures identify global and genuine entanglement, and are respectively associated with the projection and rejection of a given point of  the polytope on the corresponding biseparable segments. Solving the so called `inverse problem', we also discuss a way to force the system to behave in a particular form, which opens the possibility of controlling and manipulating entanglement for practical purposes.
\end{abstract}

%---------------------------------------> Section
\section{Introduction}

Entanglement is the most interesting nonclassical correlation of multipartite quantum systems. It represents an important resource in quantum computing, quantum information processing and quantum teleportation \cite{Nie10,Ben17,Cun19}. However, the characterization and quantification of multipartite entanglement is still an open question \cite{Bru02,Wal16}. Even in the three-qubit system case it has been pointed out that different forms of entanglement may be present \cite{Dur00,Aci00,Aci01,Sab08}. This can be seen from the fact that there is no unified notion of a maximally entangled state for more than two-qubits \cite{Dur00}. In fact, considering the most widely used measures \cite{Enr16}, different requirements comprise different characteristics of the nonlocal properties that a given system must satisfy to exhibit maximal entanglement. 

In this sense, entanglement measures based on the geometry of the state space or the appropriate projective space are of particular interest. For example, the geometric measure of entanglement (GME) evaluates the distance from a target state to its closest separable state in a given Hilbert space \cite{Shi95, Wei03}. Although its immediate geometric interpretation, the GME demands a non-trivial optimization so it becomes a nondeterministic polynomial (NP) problem \cite{Hil13}. Indeed, the amount of information that must be processed suggests that sooner or later it will be inevitable to resort to numerical methods \cite{Wei03}. Other examples include the entanglement of minimum bipartite entropy \cite{Pop03}, which quantifies the distance of a given state with its nearest state with no three-way entanglement, and robustness \cite{Elt14}.

With respect to the geometry of projective spaces, for permutation invariant states, the Majorana representation leads to the identification of maximal symmetric $n$-qubit states \cite{Gan12,Aul10}; it has also been reported an entanglement measure associated with  the barycenter of the Majorana constellation \cite{Gan12}. More recently, the triangle whose edges correspond to the squared bipartite concurrence of a three-qubit system has been considered as a useful tool to quantify entanglement \cite{Xie21}. It has been shown that the genuinely multipartite concurrence defined in \cite{Ma11} is exactly the square root of the shortest edge length of such a triangle, and that its perimeter is nothing else than the global entanglement measure considered in \cite{Mey02,Bre03}. Remarkably, the concurrence triangle area, computed through the Heron’s formula, is associated with a genuine entangled measure referred to as the concurrence fill \cite{Xie21}. Geometric simplices (tetrahedra) can be used to quantify GME \cite{Xie24}, a generalization of the Peres-Horodeckis criterion applies also for multiqubit, continuous-variable, and hybrid systems \cite{Jun11}, and even entropic measures derived from experimental correlations are useful to quantify tripartite entanglement \cite{Sch20}.

Of particular interest, the three qubit entanglement--polytope permits the introduction of some criteria for the characterization and detection of entanglement \cite{Saw13,Wal13,Agu15,Zha17,Qia18,Han04,Lun20,Lun23,Hig03,Enr18a}. To deepen the understanding of entanglement classification, the relationship between the entanglement types introduced in \cite{Aci00,Aci01} and some subsets of the polytope \cite{Han04,Lun20} has been analyzed. In this context, a relationship between the linear entropy of entanglement \cite{Mey02,Bre03} and the Euclidean distance to one of the vertices of the polytope has been observed \cite{Wal13,Mac18}. However, most of the characterization of the entanglement--polytope studied so far has been done qualitatively.

In this work, we study the entanglement between different parts of a tripartite qubit system from a purely geometric perspective. Our approach considers the entanglement polytope defined by the smallest eigenvalues of the reduced density matrices. We introduce a mapping from the state space to the polytope that leads to identifying some relationships between tripartite quantum states and points with clear geometric interpretation in the polytope. Then, fully separable, bi-separable and non-separable states are associated with concrete subsets (vertices, edges, facets, etc) of the polytope, providing a geometric identification of entanglement.

The most striking feature of our approach is to find that the projection of a given point of the polytope along the edges that represent biseparable states encodes a quantification of both, global and genuine entanglement.

As a byproduct, we look for a way to force the system to behave in a particular form. The cornerstone is provided by solutions to the inverse problem: given a point on the polytope in a region that characterizes very specific entanglement properties, the quantum state that satisfies such a profile must be found. We show that this approach offers interesting challenges and a much broader perspective on nonclassical correlations since it opens the possibility of controlling and manipulating entanglement for practical purposes. 

In Section \ref{qualitative} we review the convex structure of the three-qubit polytope and some entanglement properties are characterized qualitatively. A quantitative description of the three-qubit entanglement is provided in Section \ref{quantitative}, where we propose two distance-based entanglement measures, one quantifying the global entanglement and the other characterizing the three-qubit genuine entanglement. Section \ref{inverse} is addressed to solve the inverse problem. Some conclusions are given in Section \ref{conclu}.

%---------------------------------------> Section
\section{Qualitative characterization of entanglement}\label{qualitative}

Let $\mathcal{H} = \mathcal{H}_2 \otimes \mathcal{H}_2 \otimes \mathcal{H}_2$ be the Hilbert space of a three-qubit system, with $\mathcal{H}_2$ the two-dimensional space of states for a single qubit. Given a pure state $\vert \psi \rangle \in \mathcal{H}$, written in the standard form \cite{Aci00,Aci01},
\begin{equation}
\vert \psi \rangle = b_{0} \vert 000 \rangle + b_{1} e^{i \omega} \vert 100 \rangle + b_{2} \vert 101 \rangle + b_{3} \vert 110 \rangle + b_{4} \vert 111 \rangle, \quad b_{\ell} \geq 0, \quad \sum_{\ell =0}^{4} b_{\ell}^{2}=1,
\label{standard}
\end{equation}
with $0 \leq \omega \leq \pi$, we will pay attention to the vector $\vec{\lambda}_{\psi} = \left( \lambda_{1}, \lambda_{2}, \lambda_{3} \right)^{T}$, where $\lambda_k$ is the smallest eigenvalue of the density matrix $\rho_k$ associated to the $k$th qubit. 

Looping through all allowed values of $b_{\ell}$ and $\omega$ in (\ref{standard}), vector $\vec{\lambda}_{\psi} $ localizes the points of a convex polytope $\mathcal{P} \subset \mathbb{R}^3$ that encodes the degree of entanglement between the different parts of a tripartite qubit system \cite{Han04,Wal13,Lun20}. Therefore, we consider the mapping
\begin{equation}\label{Lmap}
\begin{tabular}{ll}
$\Lambda:$ & $\mathcal{H} \rightarrow \mathcal{P} \subset \mathbb{R}^{3}$\\[1ex]
 & $\ket{\psi} \mapsto \Lambda \left( \psi \right) = \vec{\lambda}_{\psi}$,
\end{tabular}
\end{equation}
together with the inequalities \cite{Hig03}: 
\begin{equation}
\lambda_i\leq \lambda_j+\lambda_k, \quad 0\leq \lambda_i\leq \tfrac12, \quad i,j,k =1,2,3.
\label{ineq}
\end{equation}
System (\ref{ineq}) characterizes $\mathcal P$ as the geometric scenario where the marginal problem defined by the pure state $\vert \psi \rangle$ finds solutions with physical meaning \cite{Wal13}. That is, among all possible states of one-qubit, $\lambda_k$ in (\ref{ineq}) selects those that can be achieved as the reduced one-qubit density matrix $\rho_k = \operatorname{Tr}_{i,j} \left( \vert \psi \rangle \langle \psi \vert \right)$, with $k \neq i \neq j$. No state of one-qubit that is not linked to $\vert \psi \rangle$ in this way is included in $\mathcal P$. Indeed, (\ref{ineq}) is a system of triangle inequalities that completely characterize the possible reduced one-qubit states of $\vert \psi \rangle$ \cite{Hig03}.

\begin{figure}[htbp]
\begin{center}

\includegraphics[width=0.3\textwidth]{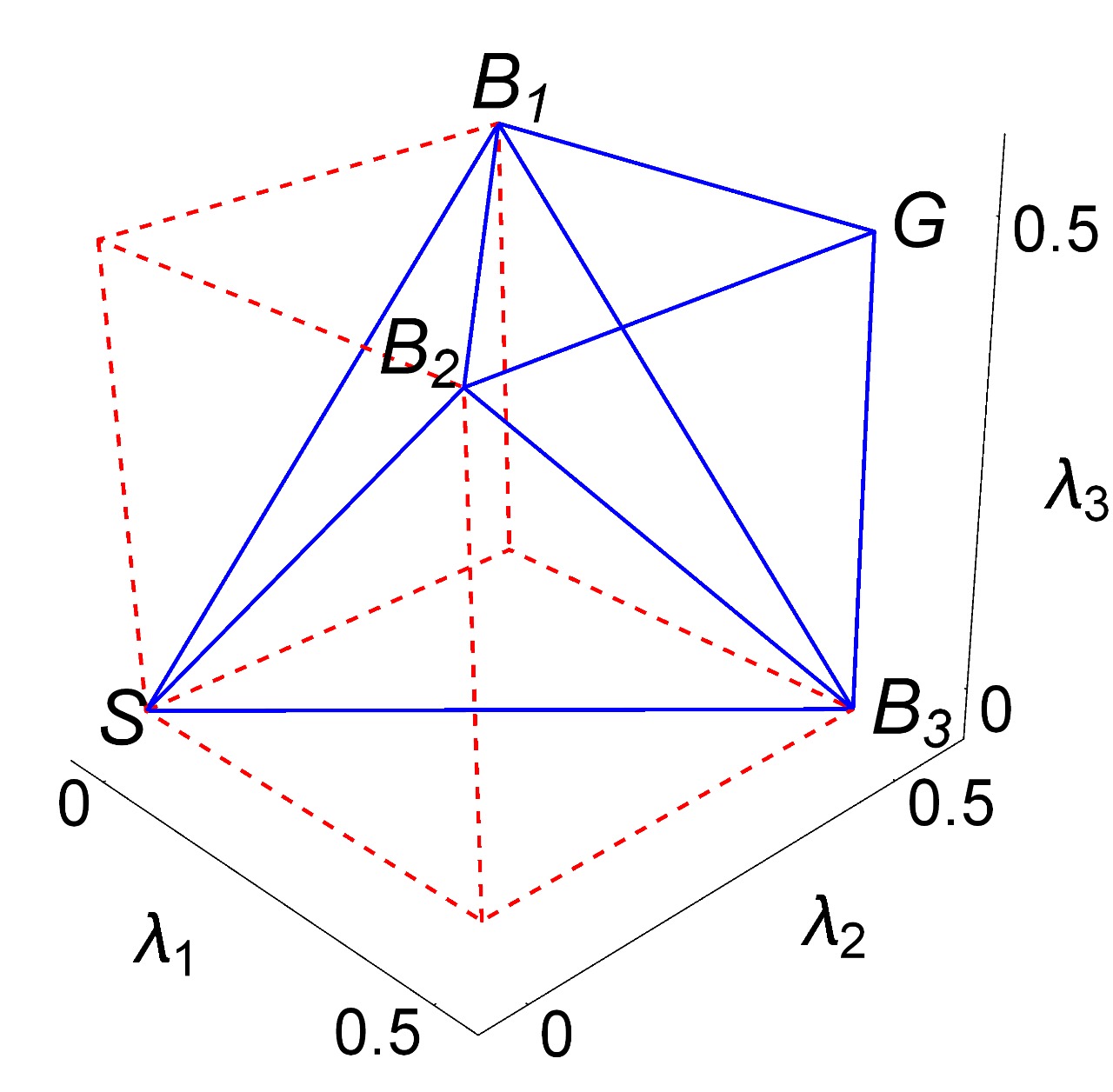} 

\caption{\footnotesize 
Entanglement--polytope $\mathcal{P} \subset \mathbb R^3$ associated with the three-qubit states (\ref{standard}). It is a body that resembles two tetrahedra, joined at their base, embedded in a cube with edges of length equal to $\sfrac12$. The parameter $\lambda_k$ (defining the corresponding axis) is the smallest eigenvalue of the density matrix associated with the $k$th qubit. Vertex $\vec S$ represents the entire set of fully separable states while $\vec G$ corresponds to the Greenberger--Horne--Zeilinger state. The vertices $\vec B_i$ correspond to bi-separable states $\vert \phi_i \rangle \vert \phi_{jk} \rangle$, with $\vert \phi_i \rangle$ the state of the $i$th qubit and $\vert \phi_{jk} \rangle$ a Bell state in the bipartition $i-jk$.
} 

\label{Poly}
\end{center}
\end{figure}

Figure~\ref{Poly} shows the entanglement--polytope $\mathcal{P}$ that we are dealing with. It is a body that resembles two tetrahedra, joined at their base, embedded in a cube with edges of length equal to $\sfrac 12$, located at the first octant of $\mathbb R^3$. To be concrete, $\mathcal{P}$ is defined by the convex combination of its vertices 
\begin{equation}\label{corners}
\begin{array}{ccc}
\vec S=(0,0,0)^T, \quad \vec B_1= \left( 0, \tfrac12 , \tfrac12 \right)^T, \quad \vec B_2 = \left( \tfrac12 , 0, \tfrac12 \right)^T,\\[1em]
\vec B_3 = \left( \tfrac12, \tfrac12, 0 \right)^T, \quad \vec G= \left( \tfrac12, \tfrac12, \tfrac12 \right)^T.
\end{array}
\end{equation}

Characterizing multipartite entanglement in terms of entanglement--polytopes is very useful since only one-particle information is required \cite{Saw13,Wal13}, see also \cite{Han04} and \cite{Enr18a}. Remarkably, this picture finds immediate application in detection processes of entanglement since its predictions do not require the measurement of correlations between the parts. For example, the experimental detection of entanglement--polytopes for three- and four-qubit genuine entanglement occurring in quantum optics has been reported in \cite{Agu15,Zha17}, a fact that shows the practical usefulness of the method. 

%%%%%%%%%%%%%%%%%%%
%%%%%%%%%%%%%%%

Considering the separability properties of $\vert \psi \rangle$, as it is written in (\ref{standard}), and remembering that local-unitary transformations preserve  entanglement properties \cite{Hor09}, one can identify different entangled pure states in terms of the coefficients $b_{\ell} \neq 0$, see details in Appendix~\ref{ApA}. Our classification coincides with the results reported in \cite{Aci00,Aci01}, and is summarized in Table~\ref{table1}.

For completeness, in Table~\ref{table1} we have included a set of states that is not obtained directly from (\ref{standard}). These states are classified as type~4d and are represented by the linear combination \cite{Car00}:
\begin{equation}
\vert \psi_{4d} \rangle = c_{0} \vert 001 \rangle + c_{1} \vert 010 \rangle + c_{2} \vert 100 \rangle + c_{3} \vert 111 \rangle, \quad c_{k} \geq 0, \quad \sum_{k=0}^3 c_k^2 =1.
\label{SII25}
\end{equation}
Note that unlike $\vert \psi \rangle$, these additional states embrace only four basis vectors. Their usefulness will become clear in the following sections.

Next, based on the results reported in \cite{Han04}, we show that this classification results in the identification of concrete convex subsets of $\mathcal P$ under the mapping $\Lambda$. 

%%%%%%%%%%%%%%%%%%%%%%%%%%%
\begin{table}[t]
\centering
\scalebox{1}{
\begin{tabular}{lll @{\vrule height 12pt depth 2pt width 0pt}}
\hline
Type & Basis product states & Subset $\subseteq \mathcal P$ \\ \hline 
1	&	$\lbrace \vert 000 \rangle \rbrace$ 		& 	 	$\vec S$ \rule[-9pt]{0pt}{12pt} \\
2a-1	& 	$\lbrace \vert 101 \rangle, \vert 110 \rangle \rbrace$ 		& 		$\overline{SB_{1}}$  \rule[-9pt]{0pt}{12pt}\\
2a-2 &	$\lbrace \vert 000 \rangle, \vert 101 \rangle \rbrace$ 		& 		$\overline{SB_{2}}$  \rule[-9pt]{0pt}{12pt}\\
2a-3 & $ \lbrace \vert 000 \rangle, \vert 110 \rangle \rbrace$ 		& 		$\overline{SB_{3}}$ \rule[-9pt]{0pt}{12pt} \\
2b & $\lbrace \vert 000 \rangle, \vert 111 \rangle \rbrace$ 		& 		$\overline{SG}$ \rule[-9pt]{0pt}{12pt} \\
3a	 &	$\lbrace \vert 000 \rangle, \vert 101 \rangle, \vert 110 \rangle \rbrace$  & 	Facets of $SB_{1}B_{2}B_{3}$ \rule[-9pt]{0pt}{12pt} \\
3b-1	& $\lbrace \vert 000 \rangle, \vert 100 \rangle, \vert 111 \rangle \rbrace$  & 	 $SB_{1}G$ \\
3b-2	& $\lbrace \vert 000 \rangle, \vert 101 \rangle, \vert 111 \rangle \rbrace$  & 	 $SB_{2}G$ \\
3b-3 & $\lbrace \vert 000 \rangle, \vert 110 \rangle, \vert 111 \rangle \rbrace$  & 	 $SB_{3}G$ \rule[-9pt]{0pt}{12pt} \\
4a	 & $\lbrace \vert 000 \rangle, \vert 100 \rangle, \vert 101 \rangle, \vert 110 \rangle \rbrace$  &  $SB_{1}B_{2}B_{3}$ \rule[-9pt]{0pt}{12pt} \\
4b-1 & $\lbrace \vert 000 \rangle, \vert 100 \rangle, \vert 110 \rangle, \vert 111 \rangle \rbrace$  & $\subset SB_{1}B_{3}G$  \\
4b-2	& $\lbrace \vert 000 \rangle, \vert 100 \rangle, \vert 101 \rangle, \vert 111 \rangle \rbrace$  & $\subset SB_{1}B_{2}G$  \rule[-9pt]{0pt}{12pt} \\
4c	 & $\lbrace \vert 000 \rangle, \vert 101 \rangle, \vert 110 \rangle, \vert 111 \rangle \rbrace$  & $SB_{1}B_{2}B_{3} \cup SB_{2}B_{3}G$ \rule[-9pt]{0pt}{12pt} \\
4d	 & $\lbrace \vert 001 \rangle, \vert 010 \rangle, \vert 100 \rangle, \vert 111 \rangle \rbrace$  &  $\mathcal{P}$ \rule[-9pt]{0pt}{12pt} \\
5 & $\lbrace \vert 000 \rangle, \vert 100 \rangle, \vert 101 \rangle, \vert 110 \rangle,\vert 111 \rangle \rbrace$  &  $\subset\mathcal{P}$  \\ \hline
\end{tabular}
}

\caption{\footnotesize 
The number of basis elements of $\mathcal{H} = \mathcal{H}_2 \otimes \mathcal{H}_2 \otimes \mathcal{H}_2$ that are included in the linear superposition (\ref{standard}) gives rise to a classification of three-qubit states \cite{Aci00,Aci01}. Some of the representative states generated by such basis elements are associated with specific convex subspaces of the entanglement--polytope $\mathcal P \subset \mathbb R^3$ shown in Figure~\ref{Poly}, see \cite{Han04} and discussion in the main text. The class of fully separable states is represented by $\vert 000 \rangle$ as a generic case, see details in Appendix~\ref{ApA}. States of type~4d are represented by the vector $\vert \psi_{4d} \rangle$ defined in (\ref{SII25}).
}
\label{table1}
\end{table}
%%%%%%%%%%%%%%%%%%%%%%%%%%%%%

The simplest class, called type~1, refers to fully separable states, hereafter written $\vert \psi_{\operatorname{sep}} \rangle = \ket{\phi_{1}} \ket{\phi_{2}} \ket{\phi_{3}}$, with $\vert \phi_{i} \rangle \in \mathcal{H}_{2}$. These states are generically represented by $\vert 000 \rangle$ in Table~\ref{table1}, since they are local-unitary equivalent to such state. Furthermore, in this case the smallest eigenvalue $\lambda_k$ of $\rho_k$ is equal to zero for all $i=1,2,3$. Thus, the entire set of vectors $\vert \psi_{\operatorname{sep}} \rangle$ is associated with the vertex $\vec S \in \mathcal{P}$, which is a 0-dimensional convex set.
 
Type~2 includes linear superpositions (\ref{standard}) with two coefficients different from zero as well as any other state $\vert \psi \rangle$ that can be transformed into any of these superpositions by means of local-unitary transformations, see Appendix~\ref{ApA} for details. We identify two different subclasses, called 2a and 2b.

Type~2a is subdivided into three different categories, called 2a-1, 2a-2 and 2a-3, whose generic states are characterized by the pairs $\{b_2, b_3\}$, $\{b_0, b_2\}$ and $\{b_0, b_3\}$, respectively. These states are bi-separable, written $\ket{\psi_{\operatorname{bi}}} = \ket{\phi_{i}} \ket{\phi_{jk}}$, with $\ket{\phi_{jk}}$ a non-separable state shared by the $j$th and $k$th qubits (the sub labels are cyclic). They correspond to the one-dimensional convex subsets of $\mathcal P$ generated by the vertices $\vec S$ and $\vec B_i$. That is, type~2a states are associated with the line-segments (edges) $\overline{SB_1}$, $\overline{SB_2}$ and $\overline{SB_3}$ of $\mathcal P$. For instance, if $\ket{\psi_{\operatorname{bi}}} = b_{2} \ket{101} + b_{3} \ket{110}$, then $\vec{\lambda}_{\psi_{\operatorname{bi}}} = (1-2\lambda) \vec{S} + 2 \lambda \vec{B}_{1}$, where $\lambda = ( 1 - \sqrt{1-4 b_{2}^{2} b_{3}^{2}} )/2$. Similarly for types~2a-2 and 2a-3. Stellar members of type~2a are the bi-separable states linked to the vertices $\vec B_i \in \mathcal{P}$, for which $\ket{\phi_{jk}}$ is one of the Bell--basis elements $\sqrt{2} \vert \Phi^{\pm} \rangle = \vert 00 \rangle \pm \vert 11 \rangle$, $\sqrt{2} \vert \Psi^{\pm} \rangle = \vert 01 \rangle \pm \vert 10 \rangle$.

The generic type~2 state is characterized by the pair $\{b_0, b_4\}$. This class of states is associated with the one-dimensional convex subset generated by the extremal points $\vec S$ and $\vec G$ (the line-segment $\overline{SG}$). Explicitly, $\vec{\lambda}_{\psi_{2b}} = (1-2\lambda) \vec{S} + 2\lambda \vec{G}$, where $\lambda = (1-\sqrt{1-4b_{0}^{2}b_{4}^{2}})/2$. The Greenberger--Horne--Zeilinger (GHZ) state 
\[
\vert GHZ \rangle = \frac{1}{\sqrt 2} \left( \vert 000 \rangle + \vert 111 \rangle \right)
\]
corresponds to the extremal point $\vec G$.

Generic states of type~3 involve three coefficients different from zero in the linear superposition (\ref{standard}). They are associated with two-dimensional convex subsets of $\mathcal P$, as we indicate below.

The type 3a refers to generic vectors that can be written in the form
\begin{equation}
\vert \psi_{\rm 3a} \rangle = \sqrt{\gamma} \vert 000 \rangle + \sqrt{\alpha} \vert 101 \rangle + \sqrt{\beta} \vert 110 \rangle, \quad 0\leq \alpha,\beta \leq 1, \quad \gamma = 1-\alpha-\beta.
\label{type3a}
\end{equation}
Therefore, the components of $\vec \lambda_{\psi_{\rm 3a}}$ are as follows
\[
\lambda_1= \tfrac12 - \left\vert \gamma - \tfrac12 \right\vert, \quad 
\lambda_2 = \tfrac12 -\left\vert  \beta - \tfrac12 \right\vert, \quad 
\lambda_3 = \tfrac12 - \left\vert \alpha - \tfrac12 \right\vert.
\]

%%%%%%%%%%%%%%%%%%%
\begin{figure}[htbp]
\begin{center}
\includegraphics[width=0.4\textwidth]{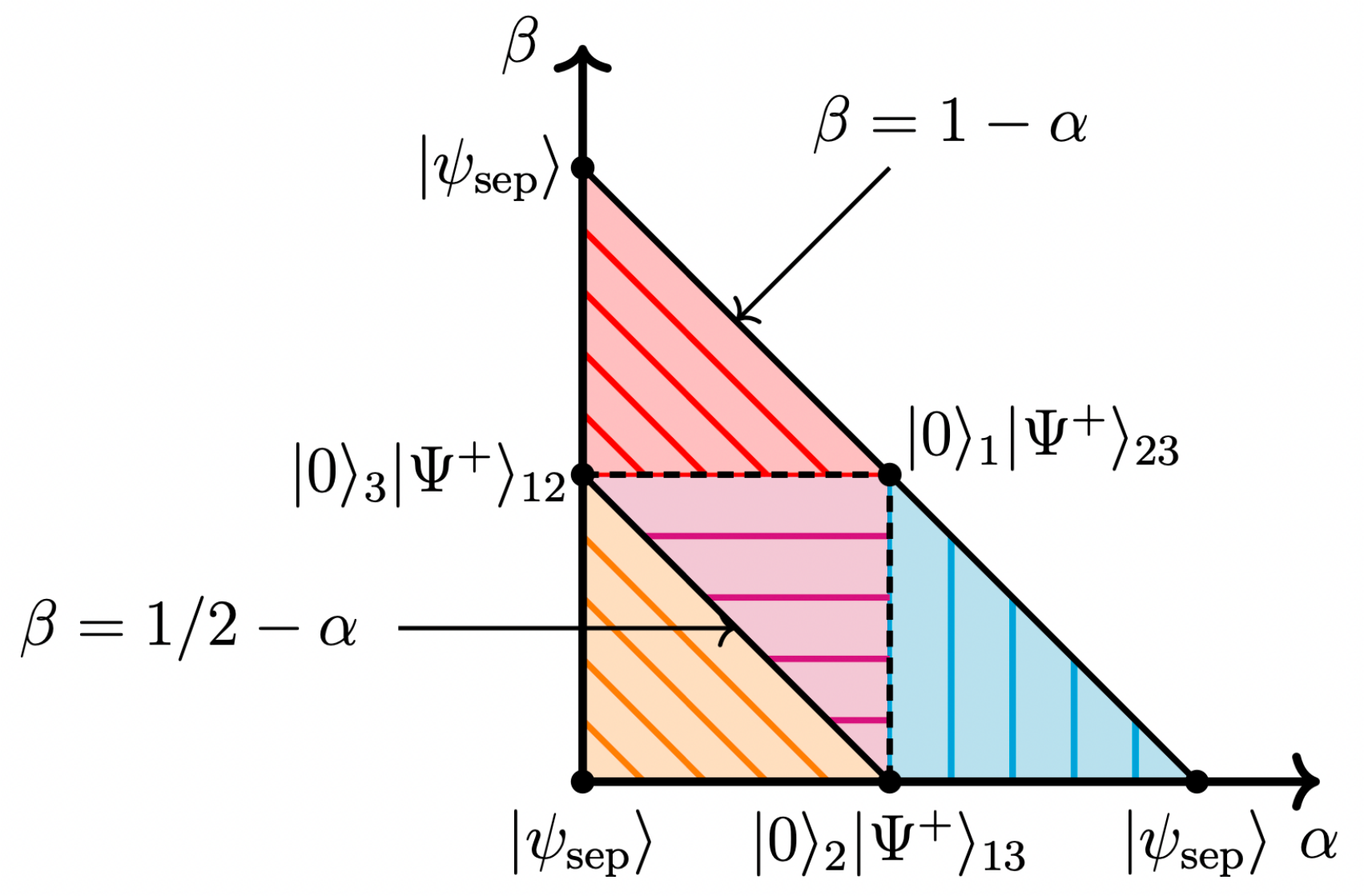}

\caption{\footnotesize 
The parameters $\alpha$ and $\beta$ completely define the state $\vert \psi_{3a} \rangle$ introduced in Eq.~(\ref{type3a}). The constraints on these parameters give rise to the convex set $\mathcal M$, a right triangle with legs of one unit on the plane $\mathbb R^2$. The extreme points, $(\alpha, \beta) = (0,0)$, $(0,1)$ and $(1,0)$, characterize $\vert \psi_{3a} \rangle$ as a fully separable state. The middle points, $(\sfrac12, \sfrac12)$, $(\sfrac12,0)$ and $(0, \sfrac12)$, produce the bi-separable states $\vert 0 \rangle_i \vert \Psi^+ \rangle_{jk}$, where $\vert \Psi^+ \rangle_{jk}$ stands for the Bell state $\vert \Psi^+ \rangle$ shared by the $j$th and $k$th qubits. The convex combinations of extreme and middle points yield four different convex subsets (regions) of $\mathcal M$ that permit a classification of entanglement for $\vert \psi_{3a} \rangle$.
}

\label{Map}
\end{center}
\end{figure}
%%%%%%%%%%%%%%%

Figure~\ref{Map} shows the convex set $\mathcal{M} \subset \mathbb R^2$ that arises from the constraints on the  parameters $\alpha$ and $\beta$ in Eq.~(\ref{type3a}). It is a right triangle with legs of one unit. The extreme points, $(\alpha, \beta) = (0,0)$, $(0,1)$ and $(1,0)$, define fully separable states that are mapped onto the vertex $\vec S$ of $\mathcal P$. In turn, the middle points, $(\sfrac12, \sfrac12)$, $(\sfrac12,0)$ and $(0, \sfrac12)$, produce the bi-separable states that are mapped onto the vertices $\vec B_i$ of $\mathcal P$. The data shown in Figure~\ref{Map} is obtained from (\ref{type3a}), after the appropriate local unitary transformation. The convex combinations of extreme and middle points yield four different convex subsets (regions) of $\mathcal M$ that permit a classification of entanglement for the state $\vert \psi_{3a} \rangle$.

\begin{figure}[htbp]
\centering
\subfloat[][Region I]{\includegraphics[width=0.2\textwidth]{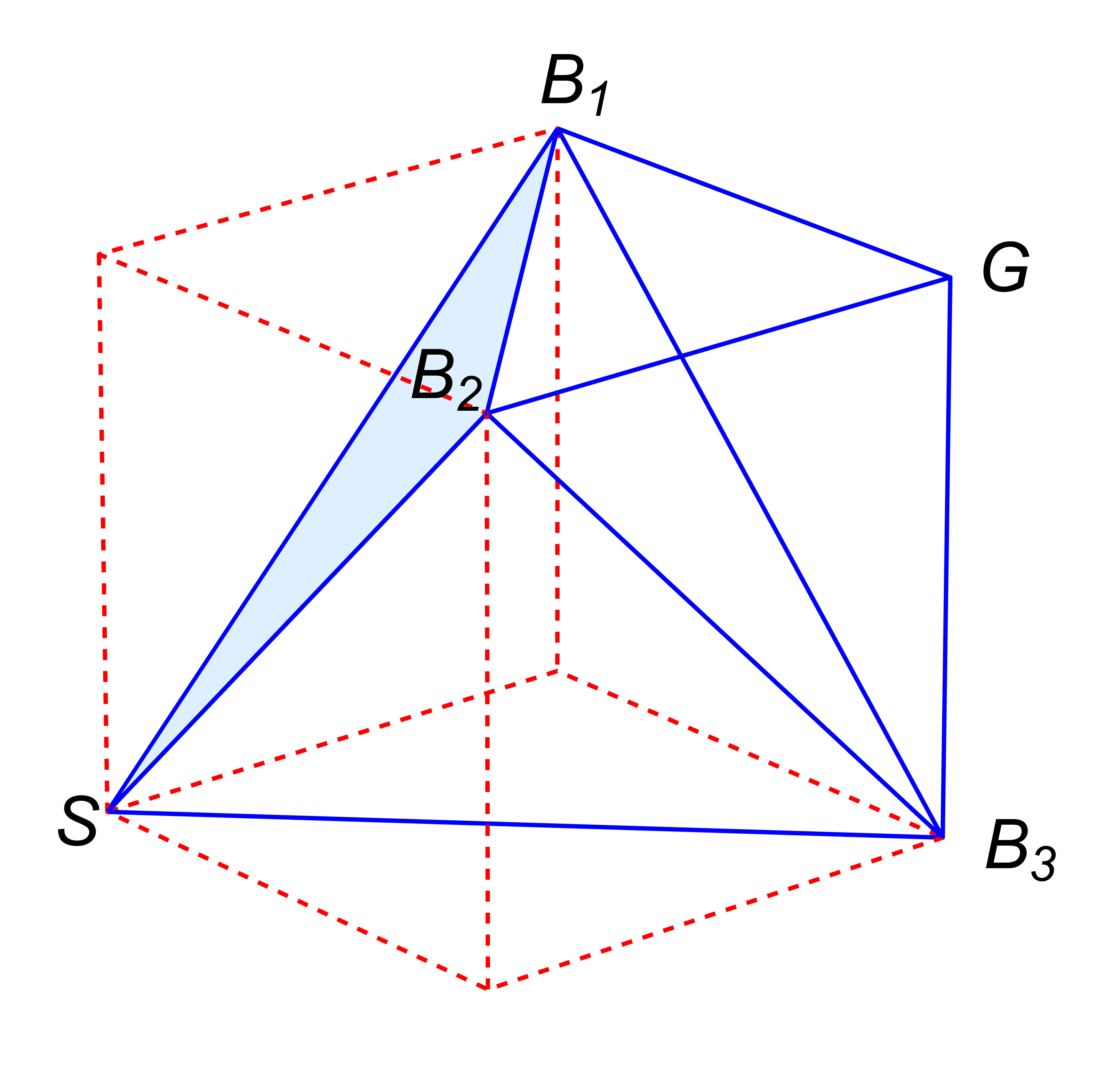}}
\hskip1cm
\subfloat[][Region II]{\includegraphics[width=0.2\textwidth]{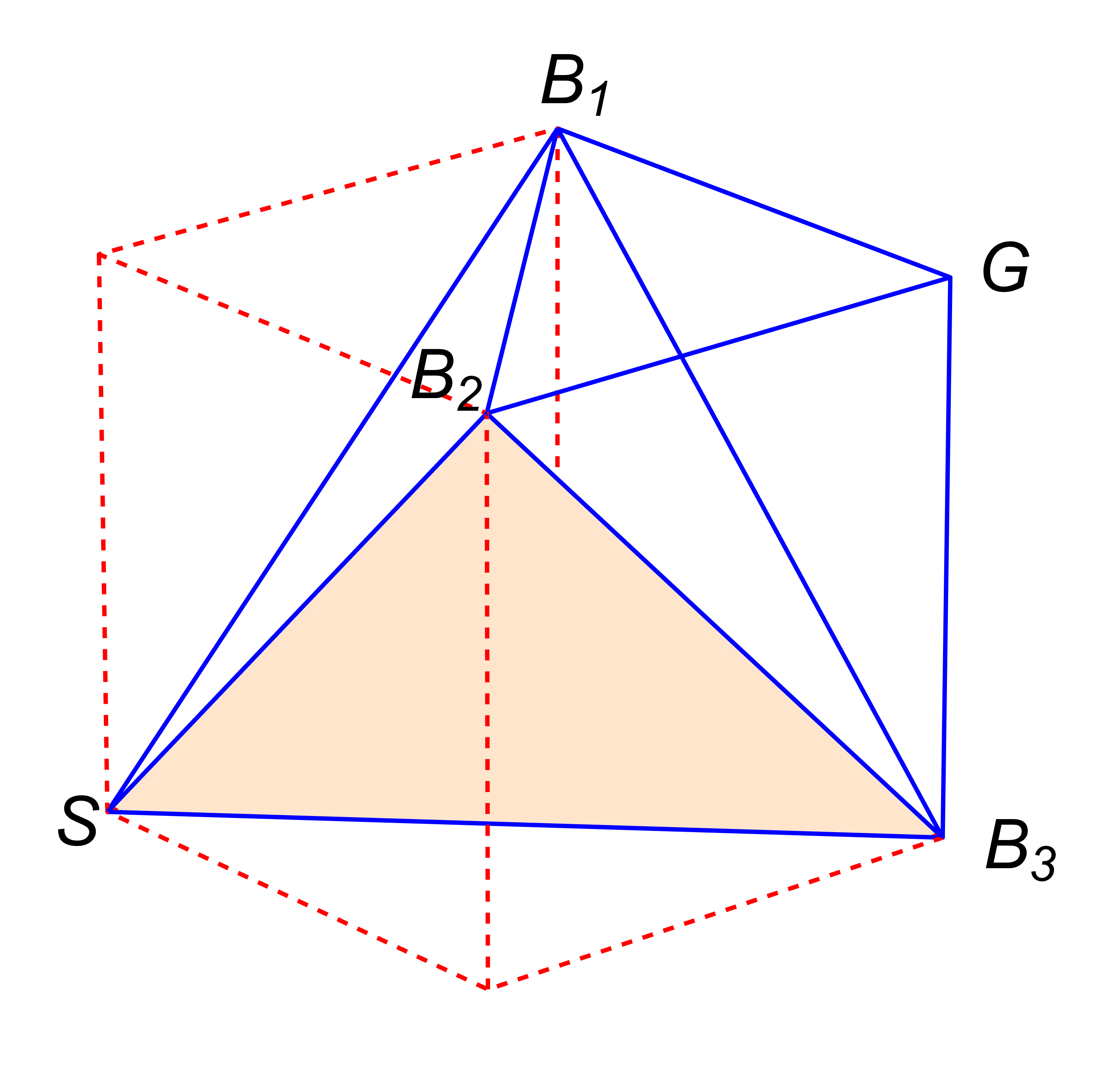}}
\hskip1cm
\subfloat[][Region III]{\includegraphics[width=0.2\textwidth]{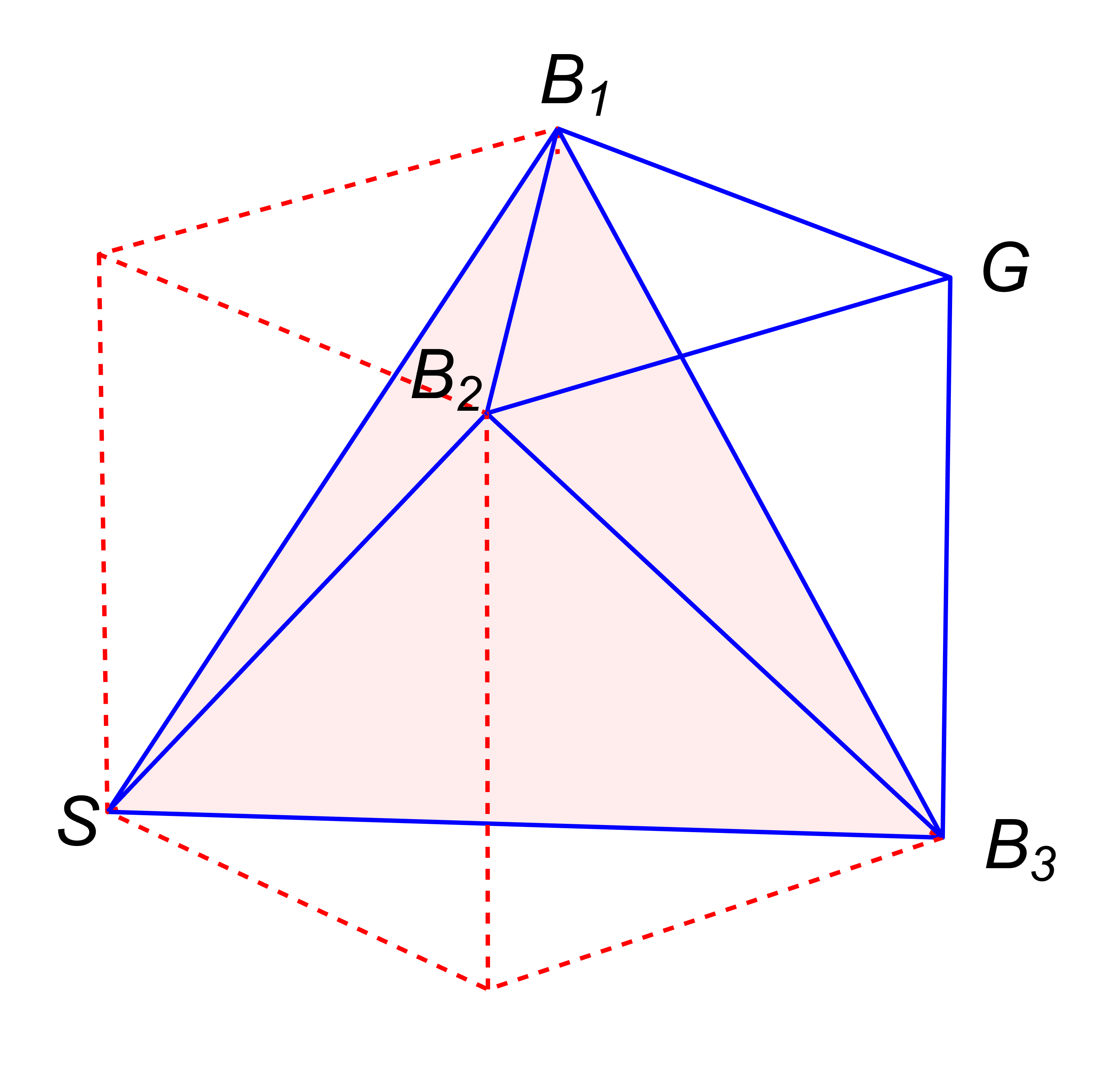}}
\hskip1cm
\subfloat[][Region IV]{\includegraphics[width=0.2\textwidth]{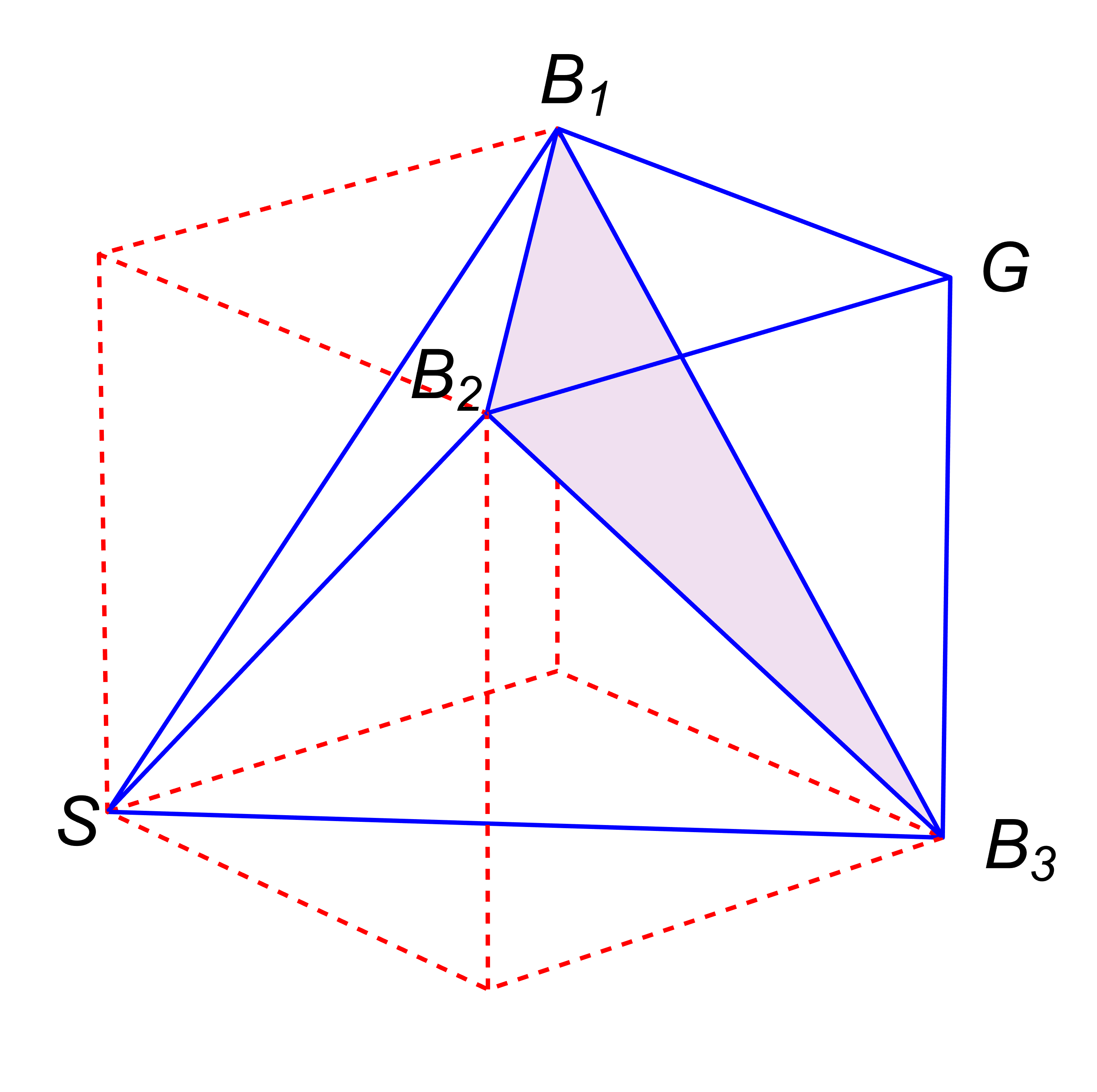}}

\caption{\footnotesize 
The four different regions of the convex set $\mathcal M$ illustrated in Figure~\ref{Map} are associated with the faces of the lower tetrahedron of the polytope $\mathcal P$. The color-code corresponds to the colors of the right-triangles in Figure~\ref{Map}.
}

\label{regions}
\end{figure}

{\bf Region I.}  The convex set $\alpha \in [\sfrac12, 1]$, $\beta \in [0, 1-\alpha]$, represented in Figure~\ref{Map} as the right triangle in blue, yields the parametrization
\begin{equation}
\vec{\lambda}_{\psi_{\rm 3a}} = \kappa_{1} \vec{S} + \kappa_{2} \vec{B}_{1} + \kappa_{3} \vec{B}_{2},
\label{eq:lambdaSB1B2}
\end{equation}
where $\kappa_{1} = 2 \alpha - 1$, $\kappa_{2} = 2 \beta $, and $\kappa_{3} = 1- \kappa_{1} - \kappa_{2}$. Vectors (\ref{eq:lambdaSB1B2}) localize the points on the triangle $\triangle_{SB_{1}B_{2}} \in \mathcal{P}$, see Figure~\ref{regions}(a).

%%%%%%%%%%%%%
%%%%%%%%%%%%%

{\bf Region II.}  The right triangle in orange of Figure~\ref{Map} corresponds to the convex set $\alpha \in [0,\sfrac12]$, $\beta \in [0, \sfrac12 - \alpha]$, and gives 
\begin{equation}
\vec{\lambda}_{\psi_{\rm 3a}} = \kappa_{1} \vec{S} + \kappa_{2} \vec{B}_{2} + \kappa_{3} \vec{B}_{3},
\label{eq:lambdaSB2B3}
\end{equation}
 with $\kappa_2=2\alpha$, $\kappa_3=2\beta$ and $\kappa_1=1-\kappa_2-\kappa_3$. In this case, the vectors (\ref{eq:lambdaSB2B3}) localize the points on the triangle $\triangle_{SB_{2}B_{3}} \in \mathcal{P}$, see Figure~\ref{regions}(b).
 
{\bf Region III.}  The right triangle in pink of Figure~\ref{Map} represents the convex set $\alpha \in [0, \sfrac12]$, $\beta \in [\sfrac12, 1 - \alpha]$. This yields 
\begin{equation}
\vec{\lambda}_{\psi_{\rm 3a}} = \kappa_{1} \vec{S} + \kappa_{2} \vec{B}_{1} + \kappa_{3} \vec{B}_{3},
\label{eq:lambdaSB1B3}
\end{equation}
 with $\kappa_1=2\alpha-1$, $\kappa_2=2\beta$ and $\kappa_3=1-\kappa_1-\kappa_2$. The points on the triangle $\triangle_{SB_{1}B_{3}} \in \mathcal{P}$ of Figure~\ref{regions}(c) are localized by the vectors (\ref{eq:lambdaSB1B3}).

{\bf Region IV.} The right triangle in light-purple of Figure~\ref{Map} stands for the convex set $\alpha \in [0, \sfrac12]$, $\beta \in [\sfrac12 - \alpha, \sfrac12]$. The corresponding vectors
\begin{equation}
\vec{\lambda}_{\psi_{\rm 3a}} = \kappa_{1} \vec{B}_1 + \kappa_{2} \vec{B}_{2} + \kappa_{3} \vec{B}_{3},
\label{eq:lambdaB1B2B3}
\end{equation}
with $\kappa_2=1-2\beta$, $\kappa_3=1-2\alpha$ and $\kappa_1=1-\kappa_2-\kappa_3$, localize the points on the triangle $\triangle_{B_{1} B_2 B_{3}} \in \mathcal{P}$ shown in Figure~\ref{regions}(d).

The latter case is of particular relevance. Making $\alpha = \beta = 1/3$ we have the state
\[
\vert \psi_{\rm 3a} \rangle = \tfrac{1}{\sqrt 3} \left( \ket{000} + \ket{101} + \ket{110} \right),
\]
which is local unitary equivalent to the well-known W--state,
\begin{equation}
\vert W \rangle = \tfrac{1}{\sqrt 3} \left( \ket{001} + \ket{010} + \ket{100} \right).
\label{S2_31}
\end{equation}
Explicitly, $\vert \psi_{\rm 3a} \rangle = \left( \sigma_{x} \otimes \mathds{1} \otimes \mathds{1} \right) \vert W \rangle$, with $\sigma_{x}$ the $x$--Pauli matrix and $\mathds{1}$ the identity operator in $\mathcal{H}_2$. Since the smallest eigenvalues are local unitary invariant \cite{Zim01,Bar01} both, $\vert \psi_{\rm 3a} \rangle$ and $\vert W \rangle$, are mapped onto the same point $ \vec{\lambda}_{W} = \left( \sfrac13,\sfrac13, \sfrac13 \right)^{T} \in \mathcal{P}$, which is the geometric center of the triangle $\triangle_{B_{1}B_{2}B_{3} }\subset \mathcal{P}$.

\begin{figure}[htbp]
\centering
\subfloat[][Type 3b-1]{\includegraphics[width=0.2\textwidth]{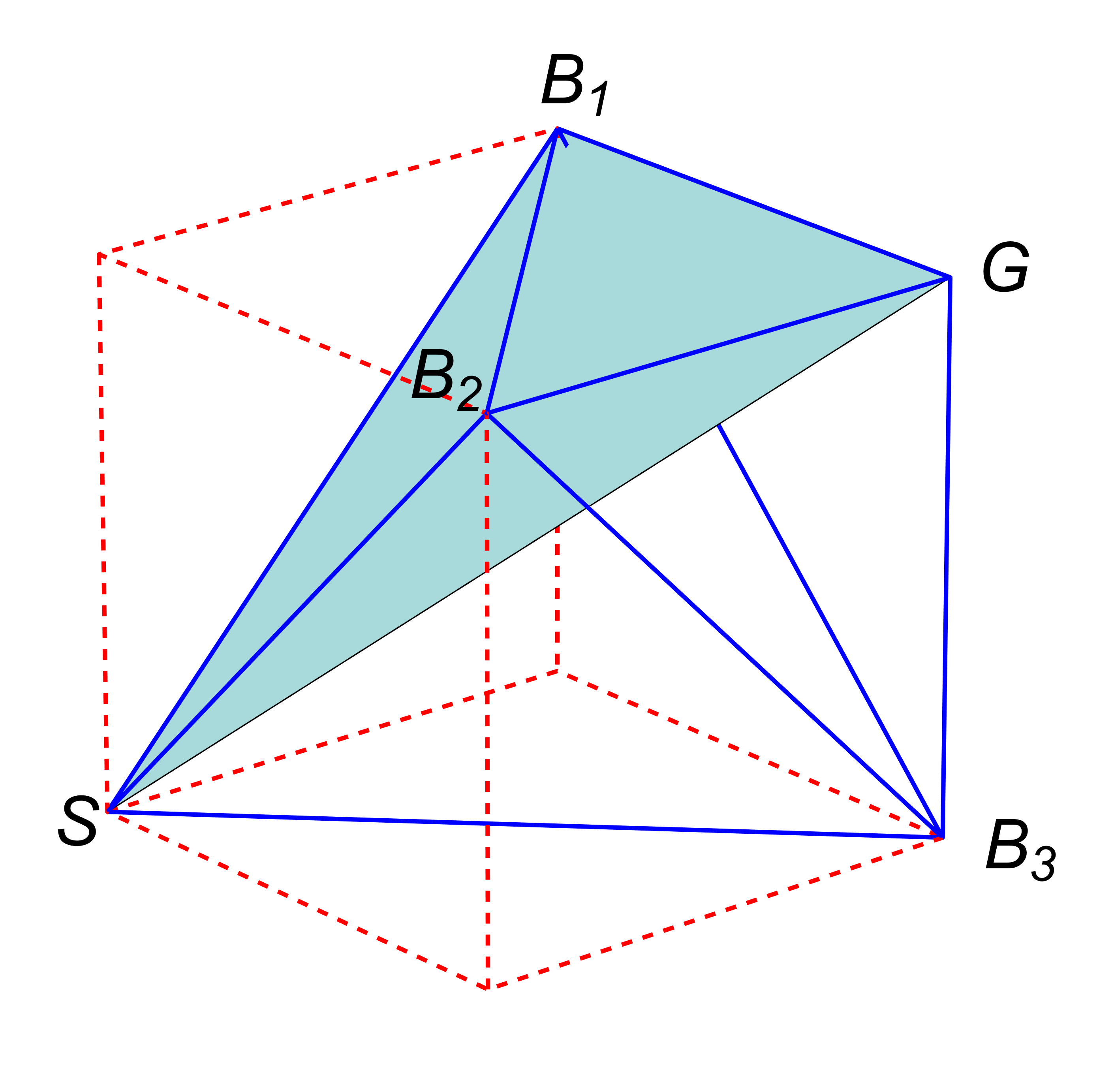}}
\hskip1cm
\subfloat[][Type 3b-2]{\includegraphics[width=0.2\textwidth]{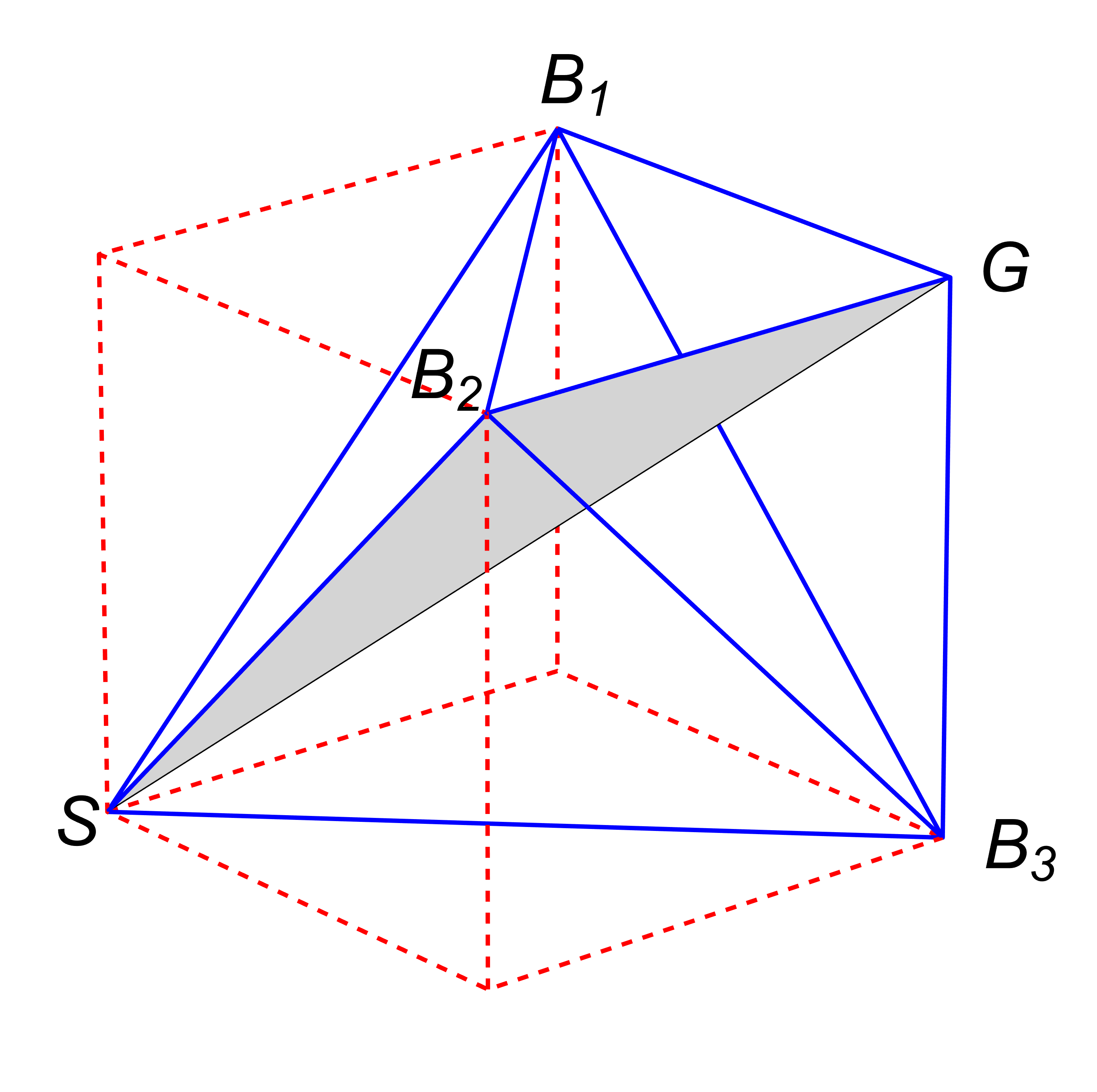}}
\hskip1cm
\subfloat[][Type 3b-3]{\includegraphics[width=0.2\textwidth]{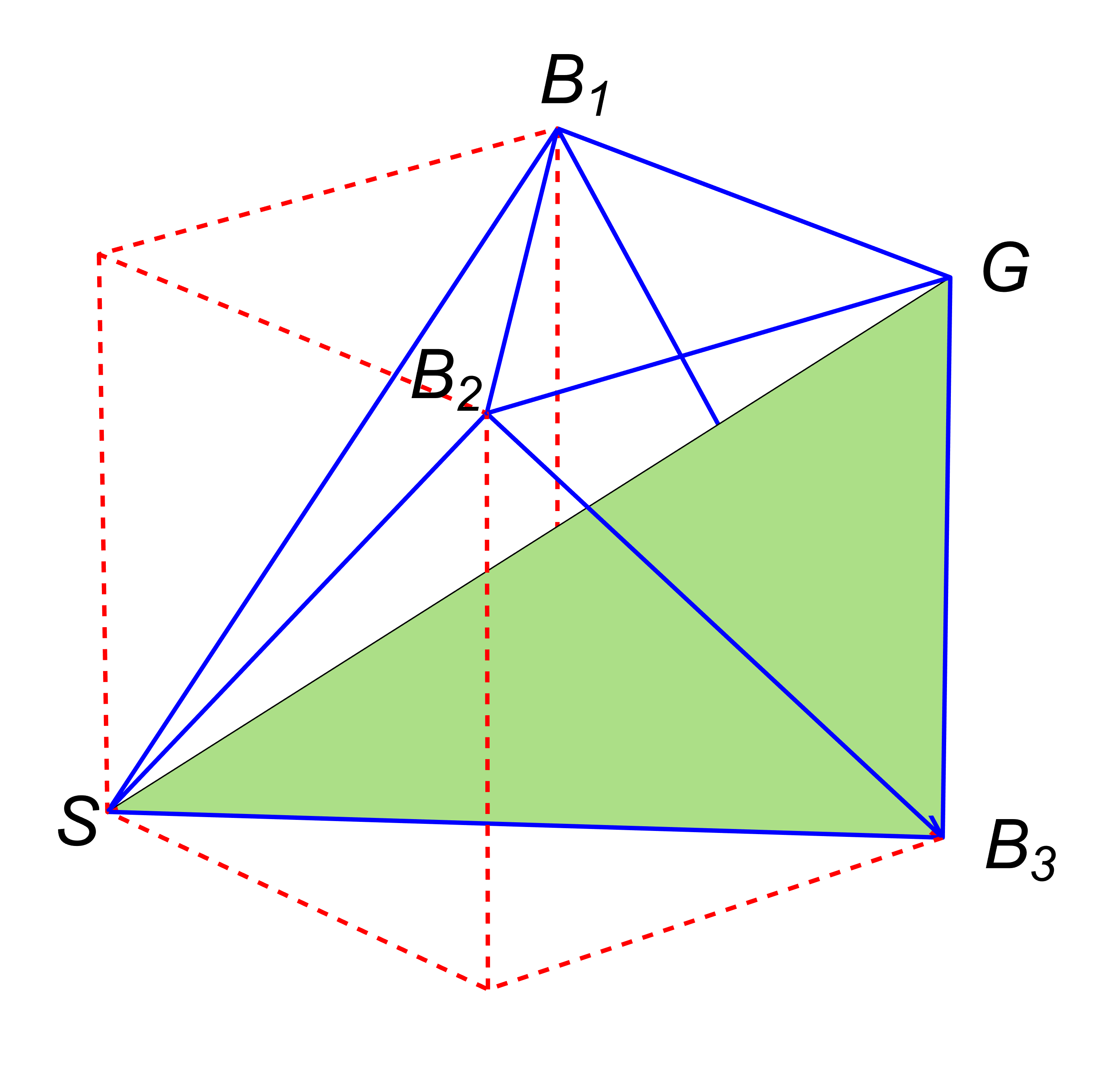}}

\caption{\footnotesize 
The type~3b-$i$ states (\ref{type3b}) are associated with the two-dimensional convex subsets that are different from the faces of the tetrahedra that form the polytope $\mathcal P$.
}

\label{t3b}
\end{figure}

On the other hand, the type 3b includes three different families of generic vectors, given by the expressions
\begin{equation}
\begin{array}{l}
\vert \psi_{\rm 3b-1} \rangle = b_{0} \vert 000 \rangle + b_{1} e^{i\omega} \vert 100 \rangle + b_{4} \vert 111 \rangle,\\[2ex]
\vert \psi_{\rm 3b-2} \rangle = b_{0} \vert 000 \rangle + b_{2} \vert 101 \rangle + b_{4} \vert 111 \rangle,\\[2ex]
\vert \psi_{\rm 3b-3} \rangle = b_{0} \vert 000 \rangle + b_{3} \vert 110 \rangle + b_{4} \vert 111 \rangle.
\end{array}
\label{type3b}
\end{equation}
They correspond respectively to the triangles $\triangle_{SB_{1}G}$, $\triangle_{SB_{2}G}$, and $\triangle_{SB_{3}G}$, see Figure~\ref{t3b}.

%%%%%%%%%%%%%
%%%%%%%%%%%%%

The generic type~4 states, referring to superpositions (\ref{standard}) that admit four non-zero coefficients, correspond to three-dimensional convex subsets of $\mathcal{P}$ that are subdivided into four different classes.

Type~4a, with $b_4=0$, is restricted to the lower tetrahedron of $\mathcal P$, written $SB_{1}B_{2}B_{3}$, where $\lambda_{1} + \lambda_{2} + \lambda_{3} \leq 1$ \cite{Han04}. The identity in this inequality is saturated by the points on the triangle $\triangle_{B_{1}B_{2}B_{3}}$.

\begin{figure}[htbp]
\centering
\subfloat[][Type 4b-1]{\includegraphics[width=0.2\textwidth]{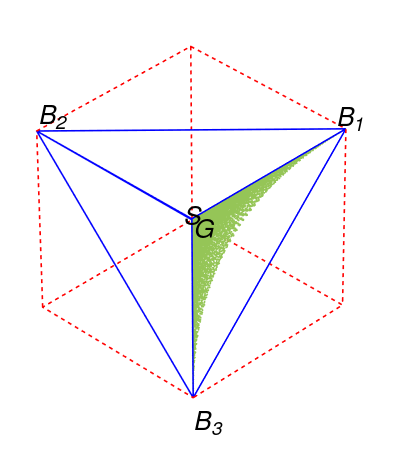}}
\hskip1cm
\subfloat[][Type 4b-2]{\includegraphics[width=0.2\textwidth]{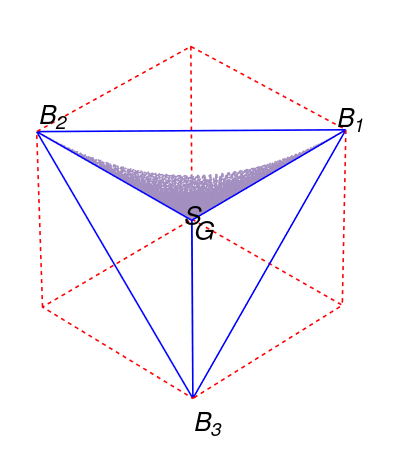}}

\caption{\footnotesize 
Type~4b states. Following \cite{Lun20}, the values of the Fourier coefficients have been discretized to identify the image of the states.
}
\label{t4b}
\end{figure}

Type~4b is subdivided into two different categories, called b-1 (with $b_2=0$) and b-2 (with $b_3=0$). They are respectively mapped into a subset of $SB_{1}B_{3}G$ and $SB_{1}B_{2}G$, see Figure~\ref{t4b}.

%%%%%%%%%%%%%
%%%%%%%%%%%%%

\begin{figure}[htbp]
\centering
\subfloat[][Type 4c-1]{\includegraphics[width=0.2\textwidth]{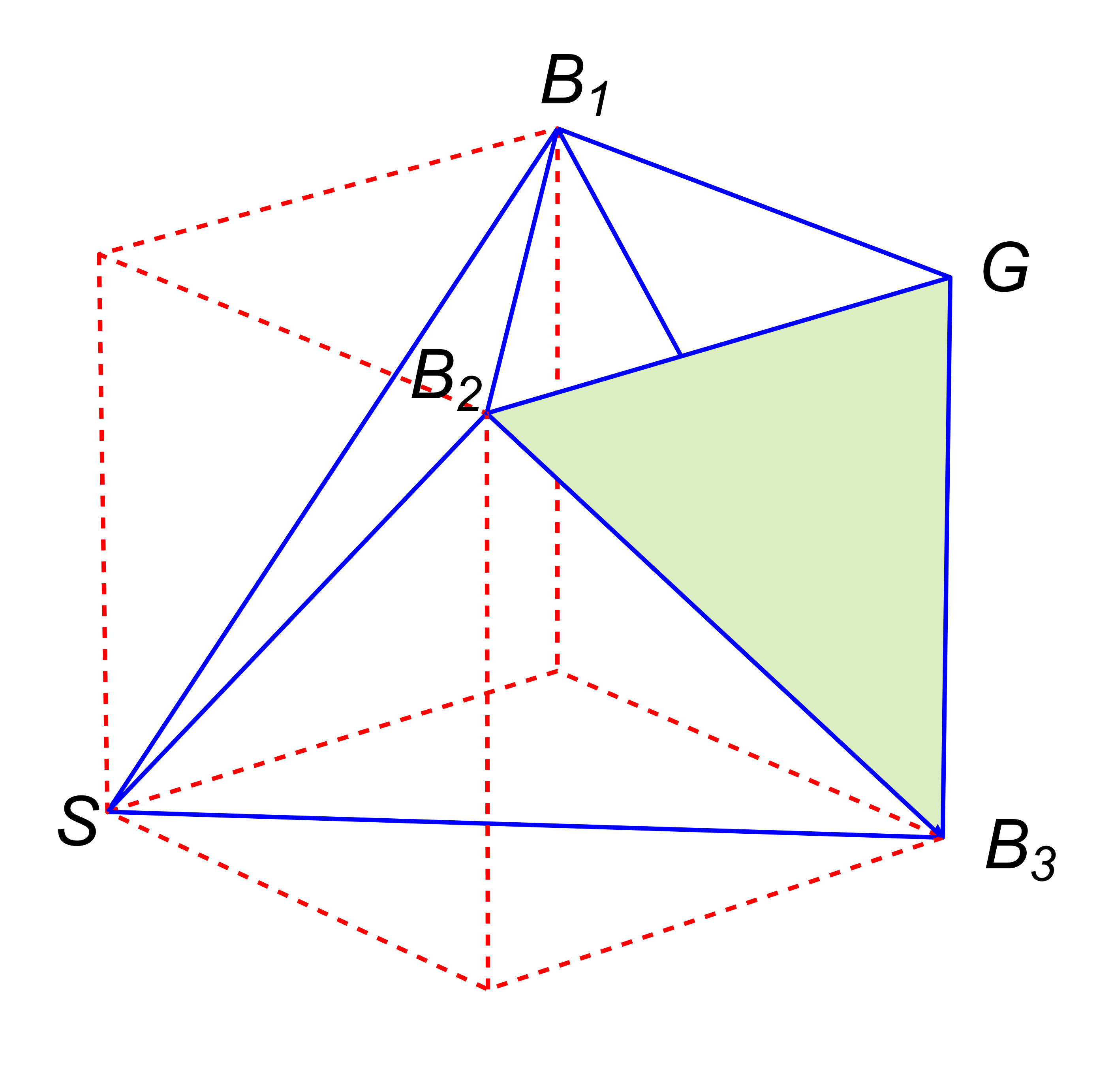}}
\hskip1cm
\subfloat[][Type 4c-2]{\includegraphics[width=0.2\textwidth]{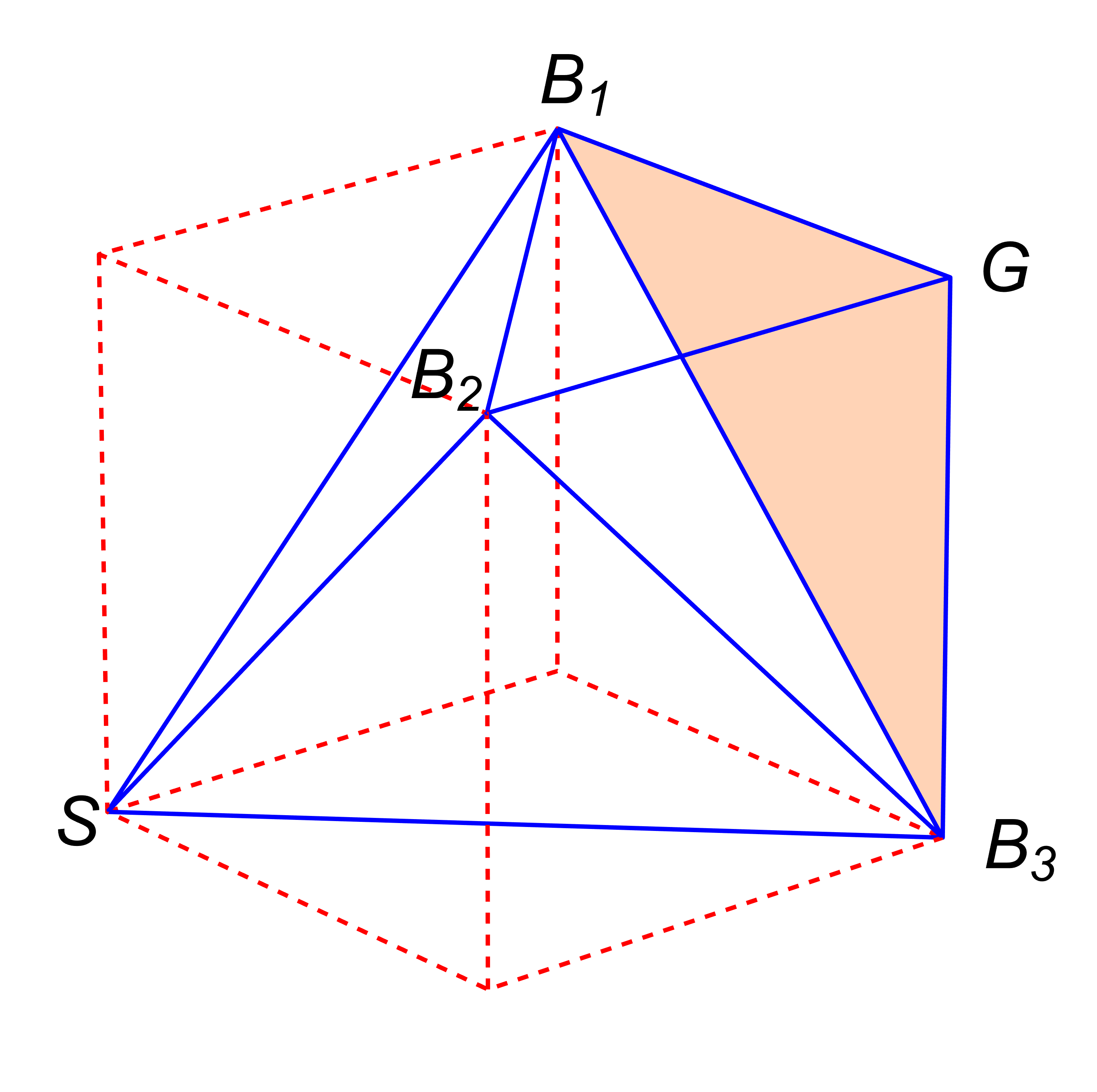}}
\hskip1cm
\subfloat[][Type 4c-3]{\includegraphics[width=0.2\textwidth]{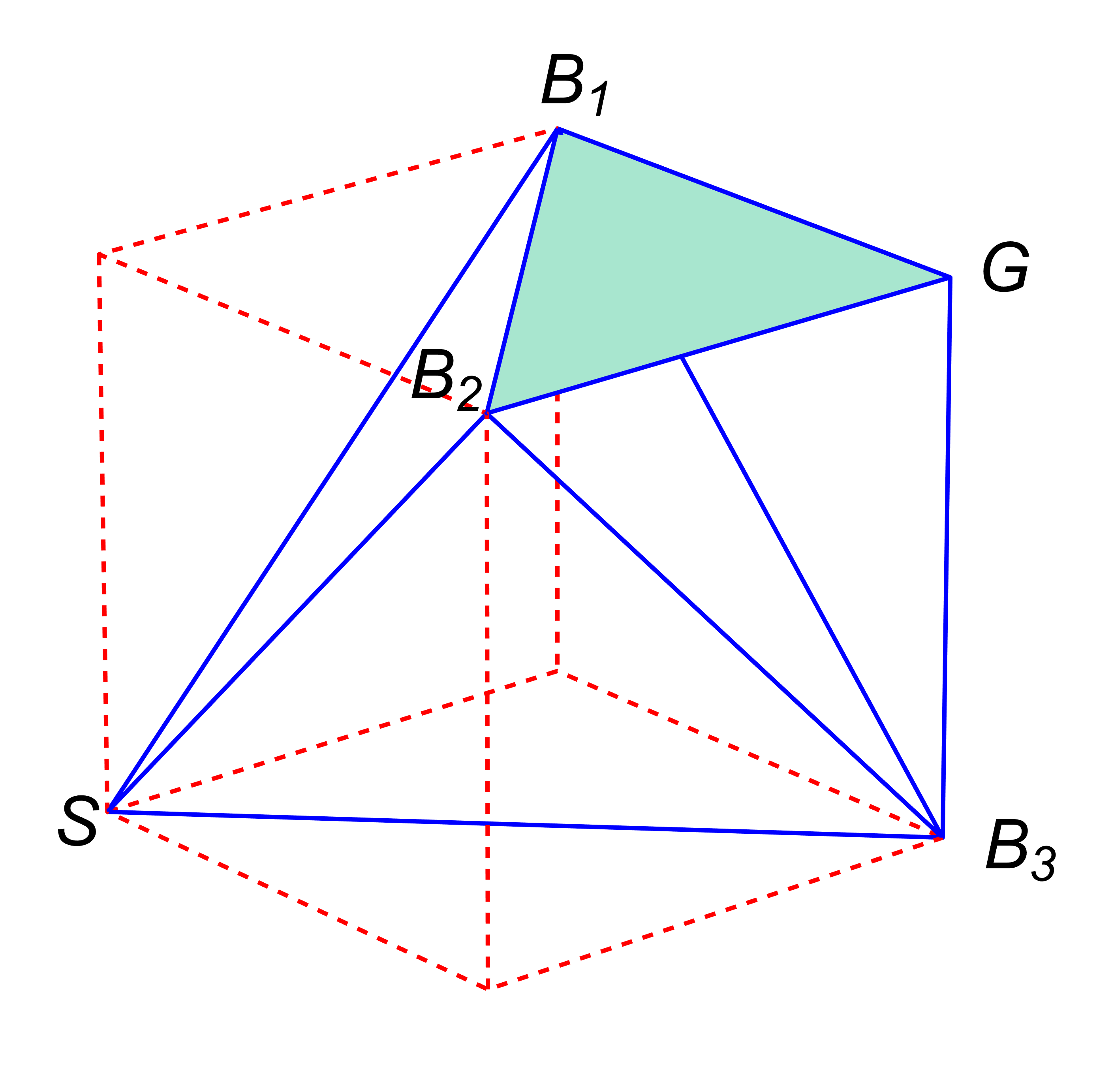}}

\caption{\footnotesize 
The type~4c-$i$ states, $i=1,2,3$, are associated with the faces of the upper tetrahedron of the polytope $\mathcal P$.
}

\label{t4c}
\end{figure}

For states of type~4c (with $b_1=0$), the image covers region $SB_{1}B_{2}B_{3} \cup SB_{2}B_{3}G$. To identify states that are mapped to the faces of the upper tetrahedron of $\mathcal P$ we fix $b_0 = 1/\sqrt 2$, and consider the linear superposition
\begin{equation}
\ket{\psi_{4c-1} }= \tfrac{1}{\sqrt{2}} \ket{000} + b_{2} \ket{101} + b_{3} \ket{110} + b_{4} \ket{111},
\label{state_4c1}
\end{equation}
where $b_{2}^{2} + b_{3}^{2} + b_{4}^{2} = \sfrac12$. In this case, the vector
\begin{equation}
\vec \lambda_{\psi_{\rm 4c-1}} = \kappa_{1} \vec{B}_{2} + \kappa_{2} \vec{B}_{3} + \kappa_{3} \vec{G},
\label{lambda_4c1}
\end{equation}
with
\begin{equation}
\kappa_{1} = 2  \sqrt{b_{2}^{2} \left( \tfrac12 - b_{3}^{2} \right)}, \quad \kappa_{2} = 2 \sqrt{b_{3}^{2} \left( \tfrac12 - b_{2}^{2} \right) }, \quad \kappa_{3} = 1 - \kappa_{1} -\kappa_{2},
\label{kappas4c1}
\end{equation}
localizes the points on the triangle $\triangle_{B_2B_3G}$. The appropriate permutations transform (\ref{state_4c1}) into a pair of additional states of type~4c. Namely,
\begin{equation}
\begin{array}{ll}
\ket{\psi_{4c-2}} = \frac{1}{\sqrt{2}} \ket{000} + b_{2} \ket{110} + b_{3} \ket{011} + b_{4} \ket{111},\\[2ex]
\ket{\psi_{4c-3}} = \frac{1}{\sqrt{2}} \ket{000} + b_{2} \ket{011} + b_{3} \ket{101} + b_{4} \ket{111},
\end{array}
\label{state_ABG}
\end{equation}
where $\sum b_{i}= \sfrac12$. It is immediate to show that these states are mapped into the triangles $\triangle_{B_1B_3G}$ and $\triangle_{B_1B_2G}$, respectively. That is, with type~4c-$i$ states, $i=1,2,3$, we cover the three faces of the upper tetrahedron in $\mathcal P$, see Figure~\ref{t4c}.

%%%%%%%%%%%%%
%%%%%%%%%%%%%

The generic type~4d state is expressed in an alternative canonical form  \cite{Car00}, given in (\ref{SII25}). This class of states is associated with points along the entire polytope $\mathcal P$. The relationship between types~4b, 4c, 4d and the polytope has been numerically studied in \cite{Lun20}.

%---------------------------------------> Section
\section{Quantitative characterization of entanglement}\label{quantitative}

Consider two different points in the entanglement--polytope, $\vec \lambda_{\psi_k} \in \mathcal{P}$, $k=1,2$. We know that there is at least one state $\vert \psi_k \rangle$ in $\mathcal H$ that is associated with each $\vec \lambda_{\psi_k}$ through the mapping $\Lambda$. 

Assuming that $\vert \psi_1 \rangle$ and $\vert \psi_2 \rangle$ have a certain degree of entanglement, what it does mean to say that one of them is more entangled than the other? 

The notion of entanglement measure gives diverse quantitative answers to such a question \cite{Ben17,Cun19,Bru02,Wal16,Sab08,Enr16,Lov07}. Most of them are based on the distance measures of quantum states, which quantify how close two states are (static case), or how well information has been preserved during a dynamic process (dynamic case) \cite{Nie10}. In any case, it is required a measure standard, a state (or set of states) for which the measure is equal to 1, and a state (or set of states) that provides a result equal to 0. For three-qubit entanglement, it is usual to associate $\vert GHZ \rangle$ with measure 1, and the fully separable states with measure 0. Within this standard, any entanglement measure of the states included in Table~\ref{table1} should range between 0 and 1.

How do the entanglement properties of $\vert \psi_k \rangle$ affect point $\vec \lambda_{\psi_k}$? 

Suppose $\vert \psi_1 \rangle$ is the GHZ--state and that $\vert \psi_2 \rangle$ is to be determined. Since $\vert GHZ \rangle$ corresponds to the vertex $\vec G = \vec \lambda_{\psi_1}$, if we wanted $\vert \psi_2 \rangle$ to be as entangled as possible, we would look for the point $\vec \lambda_{\psi_2}$ to be in the vicinity of $\vec G$. The closer $\vec \lambda_{\psi_2}$ is to $\vec G$, the state $\vert \psi_2 \rangle$ will be more `similar' to $\vert GHZ \rangle$. 

The problem is to define the notion of proximity between two points in $\mathcal P$ that corresponds to some entanglement measure in $\mathcal H$. 

Next, we provide two different options, one addressed to quantify global entanglement and the other characterizing genuine entanglement.

Global entanglement is usually quantified in terms of bipartite entanglement measures \cite{Elt14,Mey02,Hor09}. An example is the measure $Q$ introduced in \cite{Mey02}, which is the sum of concurrences between a single qubit and the remaining qubits in the system. However, $Q$ does not distinguish between fully inseparable states and entangled states that are separable according to some set of subsystems \cite{Lov07}. On the other hand, one says that a pure state is genuinely entangled if it cannot be written as the product of simpler entangled states \cite{Ma11}, the canonical examples are states $\vert GHZ \rangle$ and $\vert W \rangle$. In general, a given measure quantifies genuine entanglement if (a) it returns the result zero for fully separable states as well as for bi-separable states and (2) it is positive for non-biseparable states \cite{Ma11,Xie21,Xie24}. 

\begin{figure}[htbp]
\begin{center}
\includegraphics[width=0.15\textwidth]{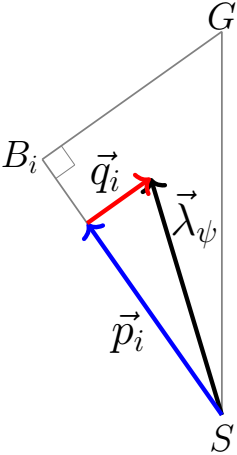}

\caption{\footnotesize 
Partial section of  the entanglement--polytope $\mathcal P$ along the plane $\Pi_i$ defined by the vertices $\vec S$, $\vec B_i$ and $\vec G$. The orthogonal projection of $\vec \lambda_{\psi} \in \mathcal{P}$ on the line-segment $\overline{SB_i} \subset \mathcal{P}$ (written $\vec p_i$), as well as the rejection of $\vec \lambda_{\psi}$ from $\overline{SB_i}$ (denoted $\vec q_i$) are useful for introducing diverse entanglement measures (see discussion in the main text). If $\vec \lambda_{\psi}$ is located in $\Pi_i$ (as shown in the figure), then $\vec q_i$ is also located in $\Pi_i$. In general, both $\vec q_i$ and $\vec \lambda_{\psi}$ localize the same (arbitrary) point of $\mathcal P$, the former from $\vec p_i$ on the line-segment $\overline{SB_i}$, the latter from the vertex $\vec S$.
}

\label{projandrej}
\end{center}
\end{figure}

Let us project the vector $\vec \lambda_{\psi} \in \mathcal P$ on the line segments $\overline{SB_{i} } \subset \mathcal{P}$. In each case one has
\begin{equation}
\vec{\lambda}_{\psi} = \vec{p}_{i} + \vec{q}_{i}, \quad \vec p_i = \left( \vec \lambda_{\psi} \cdot \mathbf{e}_{B_i} \right) \mathbf{e}_{B_i} , \quad \mathbf{e}_{B_i} = \sqrt{2} \vec B_i, \quad i=1,2,3.
\label{pq}
\end{equation}
The above decomposition is illustrated in Figure~\ref{projandrej}. Vectors $\vec p_i$ and $\vec q_i$ are the projection of $\vec{\lambda}_{\psi}$ on $\vec{B}_{i}$ and rejection of $\vec{\lambda}_{\psi}$ from $\vec{B}_{i}$, respectively. Their norms, written in terms of the components of $\vec \lambda_{\psi}$, read as follows
\begin{equation}
p_{i} = \tfrac{1}{\sqrt 2}  \left( \lambda_{j} + \lambda_{k} \right)
, \quad q_{i}=  \sqrt{\lambda_{i}^{2} + \tfrac12 \left( \lambda_{j} - \lambda_{k} \right)^{2} }, \quad  i,j,k =1,2,3.
\label{projection_rejections_norm}
\end{equation}

%%%%%%%%%%%%%
%%%%%%%%%%%%%

Clearly $\vec \lambda_{\psi} = \vec S$ produces $p_i =  q_i =  0$, since $\lambda_i=0$ for all $i=1,2,3$. Avoiding this simple case, the identity $\vec \lambda_{\psi} = \vec B_i$ for some $i=1,2,3$, provides null rejection $q_i =  0$, and vice versa. We have seen already that these points, distributed along the line segment $\overline{SB_{i} }$, yield the bi-separable states $\vert \psi_{\operatorname{bi}} \rangle = \ket{\phi_{i}} \ket{\phi_{jk}}$.

On the other hand, if $\lambda_i = \lambda$ for all $i=1,2,3$, then $p_i = \sqrt{2} \lambda$ and $q_i = \lambda$. Vectors $\vec \lambda_{\psi}$ satisfying this condition localize points along the line-segment $\overline{SG}$. In this case, the largest rejection $q_i = \sfrac12$ and largest projection $p_i= \sfrac{1}{\sqrt 2}$ are attributed to the vertex $\vec \lambda_{\psi} = \vec G$, see Figure~\ref{projandrej}.

It is notable that the norm of the projection $\vec p_i$ is proportional to the average (arithmetic mean) of $\lambda_j$ and $\lambda_k$, we write $p_i = \sqrt{2} \operatorname{Av} (\lambda_j, \lambda_k)$. Although the eigenvalue $\lambda_i$ seems to be missing here, it plays a significant role to define the decomposition (\ref{pq}). Indeed, $\lambda_i$ determines the length of $q_i$ (and the length of $p_i$ as well).

If we make $\lambda_i =0$, the inequalities (\ref{ineq}) imply $\lambda_j = \lambda_k \equiv \lambda$ (by necessity), so that $q_i=0$. That is, as noted above, $\lambda_i =0$ produces the bi-separability of the three-qubit state by isolating the $i$th qubit. Moreover, if $\lambda_i =0$ then $p_i = \sqrt{2} \lambda$. Therefore $p_i \in [0, \sfrac{1}{\sqrt 2} ]$, as expected for the bi-separable states $\vert \psi_{\operatorname{bi}} \rangle = \ket{\phi_{i}} \ket{\phi_{jk}}$. 

In addition, since $\lambda_i = \sfrac12$ means that $\rho_{i}$ is maximally mixed, the corresponding three-qubit state (in the bipartition $i-jk$) is maximally entangled. This configuration imposes the number $\sfrac{1}{(2 \sqrt 2)}$ as a lower bound on the values of $p_i$, for which we find maximally entangled states. As a consequence, the projection domain is restricted to the second half of the previous case, $p_i \in [\sfrac{1}{(2 \sqrt 2)}, \sfrac{1}{\sqrt 2}]$. 

The previous analysis shows that the norm of $\vec p_i$ encodes sensitive information about the entanglement of the three-qubit system.

Looking for a global treatment, where all the components of $\vec \lambda_{\psi}$ intervene at the same time, we propose the following arithmetic mean
\begin{equation} 
\label{measure_xi}
\xi (\psi) 
= \tfrac{\sqrt 2}{3} \left( p_1 + p_2 + p_3 \right) 
= \tfrac23 \left( \lambda_1 + \lambda_2 + \lambda_3 \right),
\end{equation}
which vanishes for fully separable states ($\lambda_{1,2,3}=0$), and is normalized to 1 for the GHZ--state ($\lambda_{1,2,3}=\sfrac12$).

%%%%%%%%%%%%%
\begin{figure}[htbp]
\centering
\subfloat[][]{\includegraphics[height=0.3\textwidth]{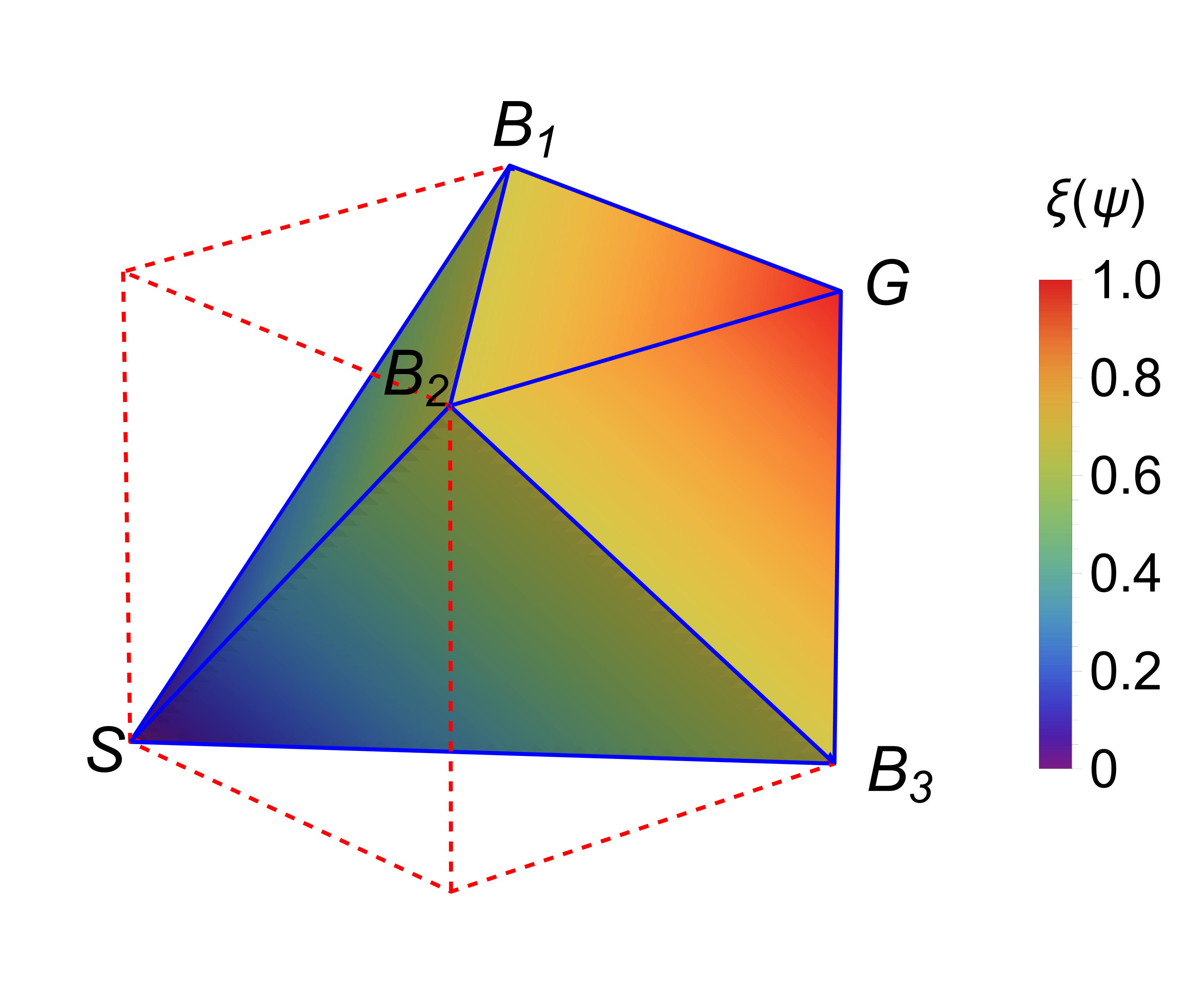}}
\hskip2cm
\subfloat[][]{\includegraphics[height=0.3\textwidth]{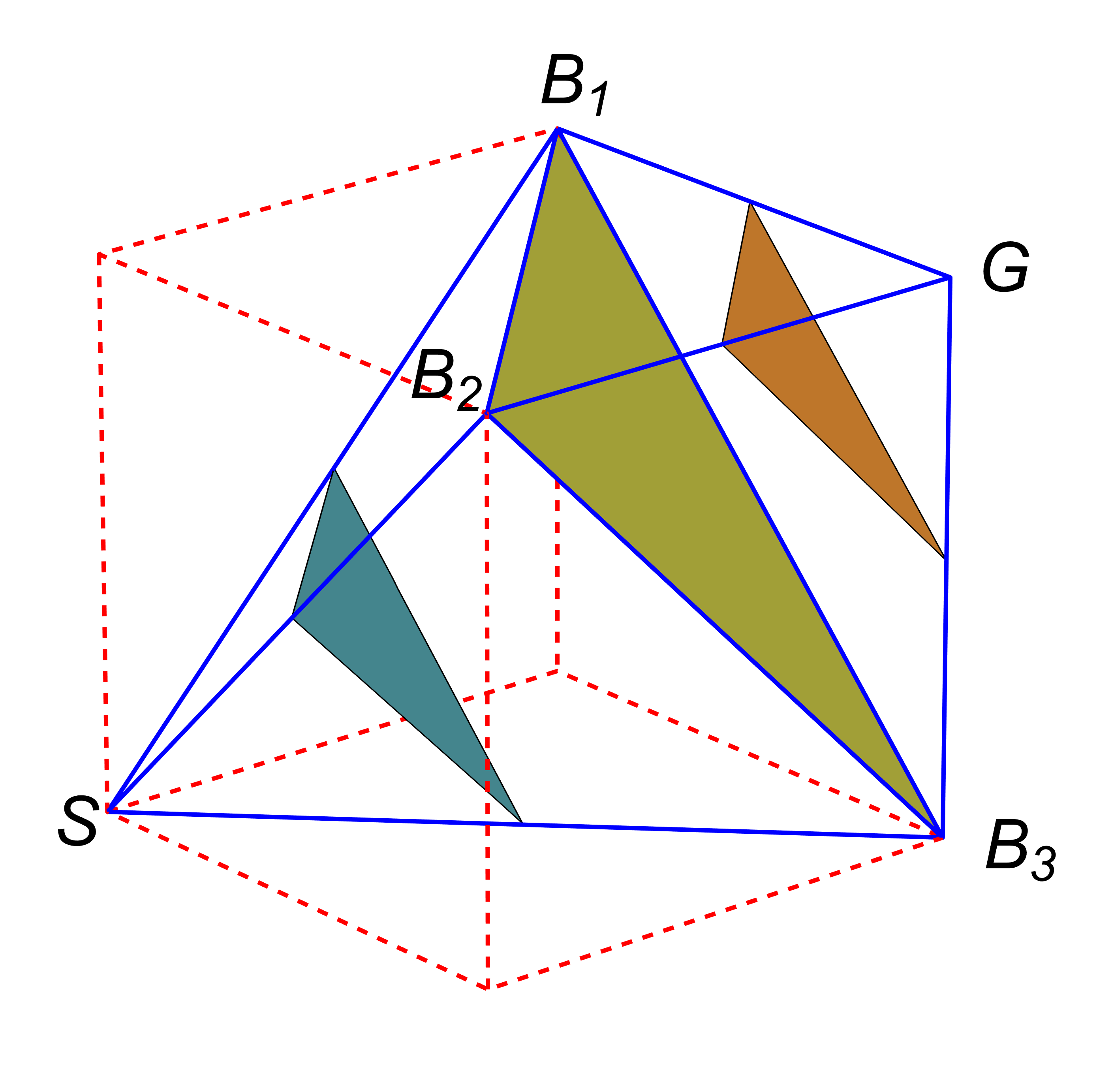}}

\caption{\footnotesize 
The arithmetic mean $\xi (\psi)$ introduced in Eq.~(\ref{measure_xi}) vanishes for fully separable states and is normalized to 1 for the GHZ--state. (\textbf{a}) The points $\vec \lambda_{\psi} \in \mathcal P$ average higher values $\xi$ as they get closer to the vertex $\vec G \in \mathcal P$. (\textbf{b}) Condition $\xi (\psi) = \xi_0$, with $\xi_0 =
\operatorname{constant}$, defines planes transverse to the line segment $\overline{SG}$ whose points represent states $\vert \psi \rangle \in \mathcal H$ with the same degree of entanglement. The $\xi_0$--planes shown in the figure correspond to $\xi_0 = \sfrac13, \sfrac23$, and $\sfrac56$.
}

\label{xi}
\end{figure}
%%%%%%%%%%%%%

Figure~\ref{xi}(a) shows the distribution of points $\vec \lambda_{\psi}$ according to $\xi (\psi)$. Clearly, these points average better values $\xi$ as they get closer to the vertex $\vec G$. 

Of particular interest, if $\lambda_i = \lambda$ for all $i=1,2,3$, then $p_i = \sqrt{2} \lambda$, $q_i = \lambda$, and $\xi (\psi) = 2 \lambda$. In this case $\vec \lambda_{\psi}$ localizes points along the line-segment $\overline{SG}$. Then the projection $p_i$ ($=\sfrac{1}{\sqrt 2}$), the rejection $q_i$ ($=\sfrac12$) and the measure $\xi$ ($=1$) reach their maximum when $\vec \lambda_{\psi} = \vec G$.

A very versatile property of the average (\ref{measure_xi}) arises by imposing the condition $\xi (\psi) = \operatorname{constant}$, since this defines a plane transverse to the line segment $\overline{SG}$ whose points represent states $\vert \psi \rangle \in \mathcal H$ with the same degree of entanglement. Indeed, assume that such a plane cuts the line segment $\overline{SG}$ at the point localized by $\vec \lambda_0$. We write $\vec \lambda_0 = \xi_0 \vec G$, with $\xi_0\in[0,1]$. If $\vec \lambda$ localizes another point on the plane, then $\vec \lambda - \vec \lambda_0$ must be orthogonal to the unitary vector $\mathbf{n} = \vert \vert \vec G \vert \vert^{-1} \vec G$, so we arrive at the equation
\begin{equation}\label{xiAplane}
\hat n\cdot (\vec \lambda-\vec \lambda_0)=0.
\end{equation}
The solution of (\ref{xiAplane}) is easily found, it is given by $\xi_0 = \xi (\psi)$. 

Figure~\ref{xi}(b) shows some $\xi_0$--planes. Among them, the triangle $\triangle_{B_1B_2B_3}$ coincides with the plane $\xi_0 = \sfrac23$ and contains the W--state at its geometric center, as indicated above. That is, $\vert W \rangle$ is $\xi = \sfrac13$ distant from $\vert GHZ \rangle$.

The decomposition (\ref{pq}) offers another possibility to quantify entanglement. As we have seen, the rejection $\vec q_i$ is very sensitive to the eigenvalue $\lambda_i$ (its norm $q_i$ cancels if $\lambda_i=0$), and vice versa. Then, we may use $q_i$ to quantify the `transverse distance' between $\vec \lambda_{\psi}$ and $\overline{SB_i}$ since vector $\vec \lambda_{\psi}$ is as `close' to $\overline{SB_i}$ as the rejection $q_i$ (equivalently, $\lambda_i$) approaches zero. In this context, we introduce the function
\begin{equation}\label{mu1}
 \mu (\psi) = 2 \min (q_1,q_2,q_3),
\end{equation}
which quantifies the transverse--distance from $\vec \lambda_{\psi}$ to the nearest line segment $\overline{SB_i}$.

\begin{figure}[htbp]

\centering
\subfloat[][]{\includegraphics[height=0.3\textwidth]{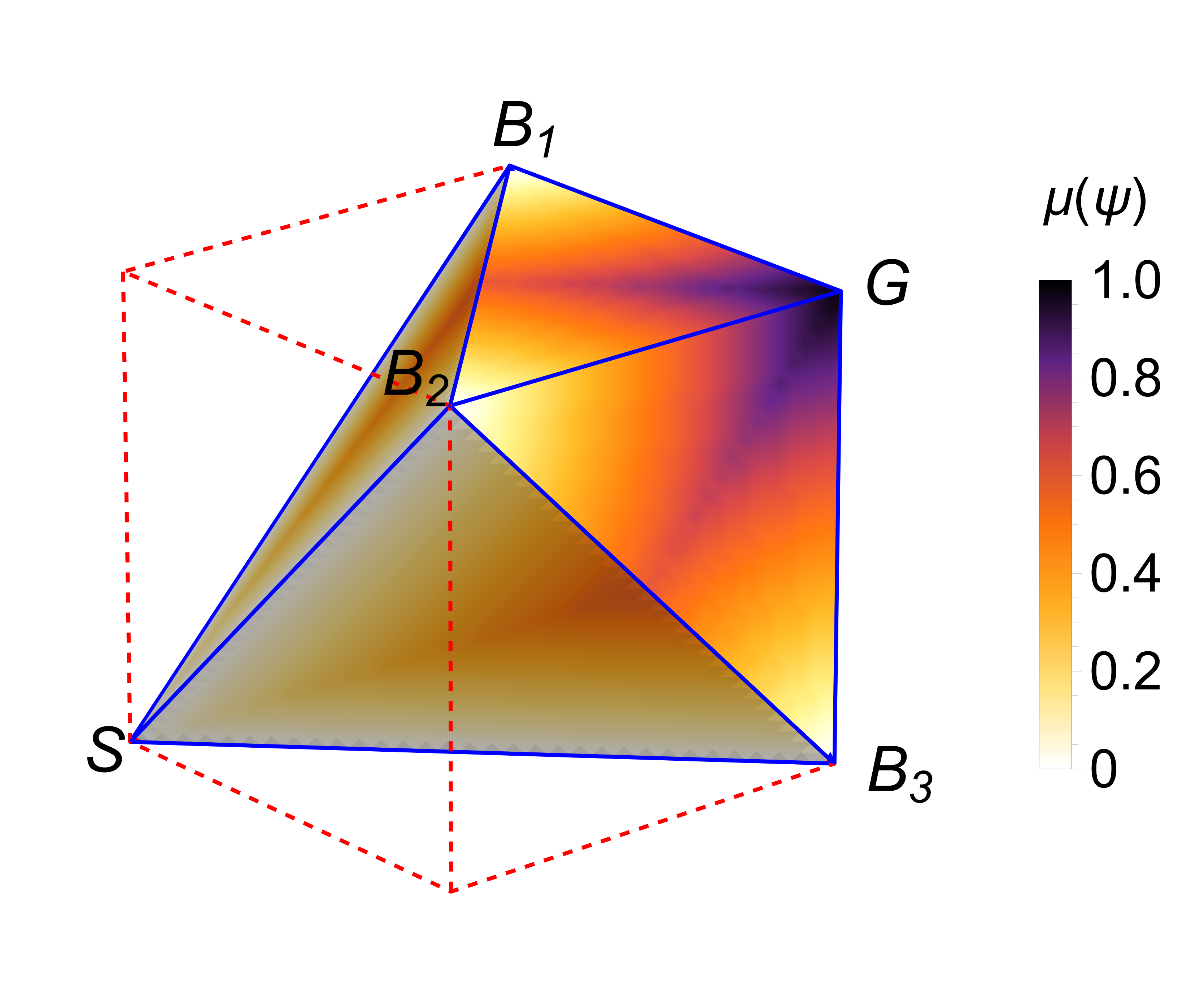}}
\hskip2cm
\subfloat[][]{\includegraphics[height=0.3\textwidth]{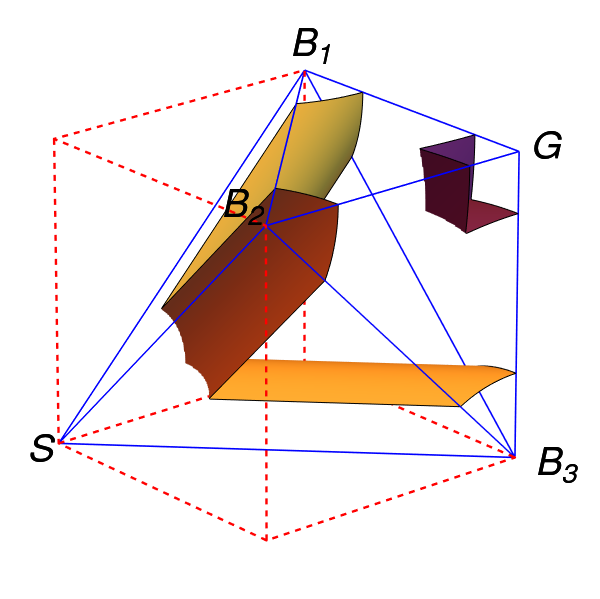}}

\caption{\footnotesize 
({\bf a}) The transverse--distance measure $\mu (\psi)$ introduced in (\ref{mu1}) gives 0 for separable states and 1 for the GHZ--state. In addition, this also gives 0 for bi-separable states, so it discriminates bi-separability from genuine entanglement. Notably, $\mu$ gives the same value as $\xi$ for the W--state, which confirms that $\vert W \rangle$ is $\xi = \mu = 1/3$ distant from $\vert GHZ \rangle$. ({\bf b}) Regions of constant measure $\mu (\psi) = \mu_0$ for $\mu_0 = 4/5$ (closest to vertex $G$) and $\mu_0 = 1 - \sqrt{13}/5$ (closest to vertex $S$).
}

\label{mutopo}
\end{figure}

Figure~\ref{mutopo}(a) shows the distribution of points $\vec \lambda_{\psi}$ according to the function $\mu$. Similar to the previous case, condition $\mu (\psi) = \mu_0 = \operatorname{const}$ provides a collection of points that represent pure states $\vert \psi \rangle$ with the same degree of entanglement. Figure~\ref{mutopo}(b) shows the regions of $\mathcal P$ associated with two different values of $\mu_0$.

%%%%%%%%%%%%%
%%%%%%%%%%%%%

\begin{figure}[htbp]
\centering
\includegraphics[height=0.2\textwidth]{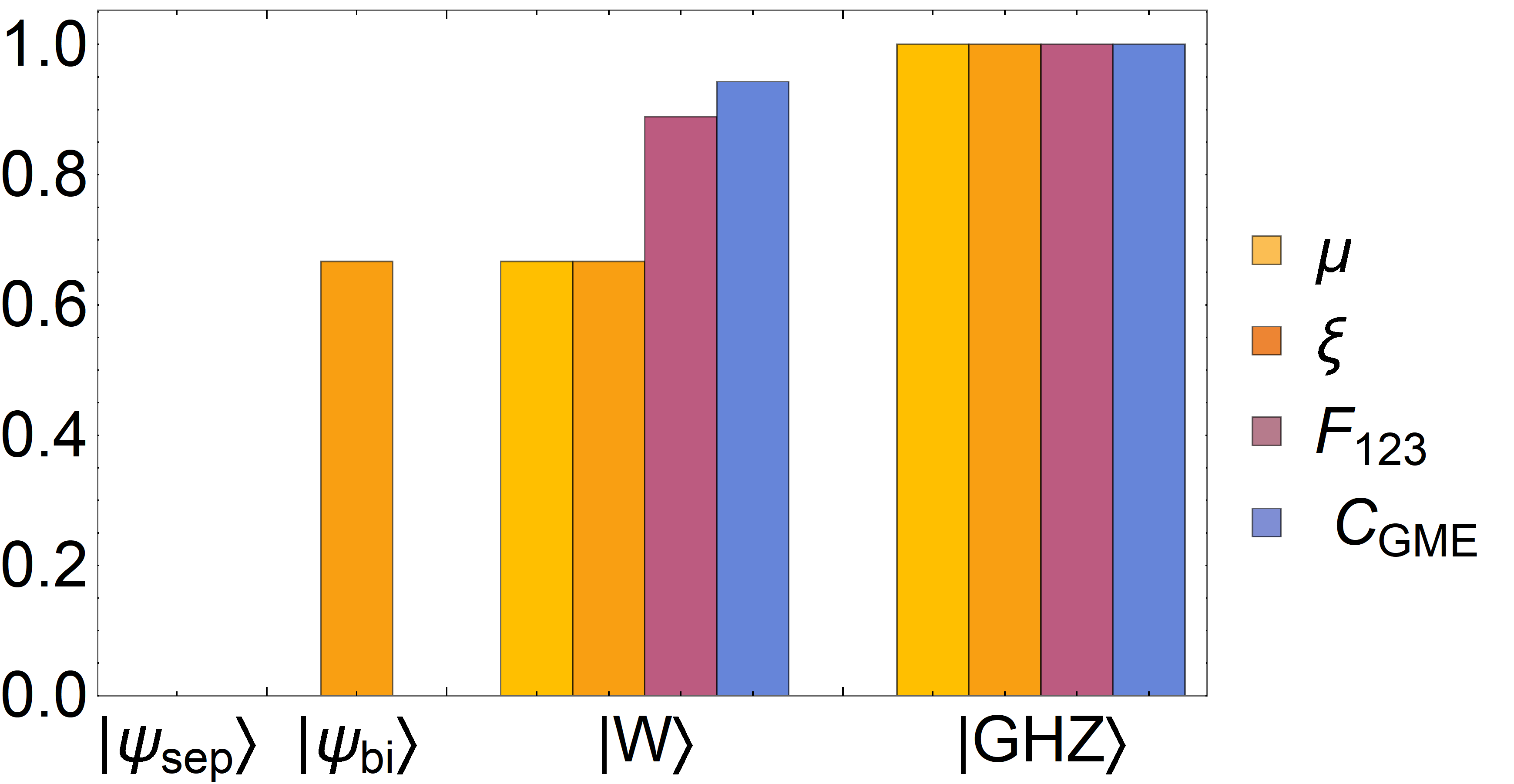}

\caption{\footnotesize
Entanglement measures $\xi (\psi)$, $\mu (\psi)$, $F_{123} (\psi)$ and $\mathcal{C}_{GME} (\psi)$ for states $\vert \psi_{\operatorname{sep}} \rangle$, $\vert \psi_{\operatorname{bi}} \rangle = \vert 0 \rangle_i \vert \Psi^+ \rangle_{jk}$, $\vert W \rangle$ and $\vert GHZ \rangle$. All the measures give 0 for fully separable states. Measure $\xi$ does not discriminate bi-separability from genuine entanglement since it averages a value different from 0 for bi-separable states. Among these measures, $\mu$ gives the lowest results for the four states.
}

\label{barras}
\end{figure}

The standard of $\mu(\psi)$ is as follows: it vanishes for separable states, attains its maximum value ($=1$) for the GHZ--state, and carries out the value 
$\mu = \sfrac23$ for the W--state. Incidentally, $\mu(\psi)$ gives the same value as $\xi(\psi)$ for the W--state, reinforcing our previous statement that $\vert W \rangle$ is $\xi = \mu = \sfrac13$ distant from $\vert GHZ \rangle$, see Figure~\ref{barras}.

As $q_i=0$ is a condition for bi-separability, in contrast with $\xi$, the measure $\mu$ also gives the value 0 for bi-separable states, see Figure~\ref{barras}. Indeed, $\mu$ is a measure of genuine multipartite entanglement since it satisfies the requirements reported in, for example, \cite{Xie21}.

%%%%%%%%%%%%%%
%%%%%%%%%%%%%%

Figure~\ref{barras} shows the values provided by the measures $\xi$ and $\mu$ for states $\vert \psi_{\operatorname{sep}} \rangle$, $\vert \psi_{\operatorname{bi}} \rangle =\vert 0 \rangle_i \vert \psi^+ \rangle_{jk}$, $\vert W \rangle$ and $\vert GHZ \rangle$. These results are compared with the values that can be obtained in terms of the concurrence--fill $F_{123}$ and the genuine multipartite concurrence $\mathcal{C} _{\operatorname{GME}}$. The former measure is defined as \cite{Xie21}:
\[
F_{123} (\psi) = \left[ 8 Q(\psi) \left( \frac{3}{2} Q(\psi) - \mathcal{C}_{1}^{2} \right) \left( \frac{3}{2} Q(\psi) - \mathcal{C}_{2}^{2} \right) \left( \frac{3}{2} Q(\psi) - \mathcal{C}_{3}^{2} \right) \right]^{1/4},
\]
where $Q \left( \psi \right) = \frac{1}{3} \sum_{i=1}^{3} \mathcal{C}_{i}^{2} \left( \psi \right)$. The $I$ concurrence $\mathcal{C}_{i}$ is a measure of entanglement between the $i$th qubit and the subsystem composed of qubits $j$ and $k$ \cite{Run01}. In turn, the measure $\mathcal C_{GME}$ is defined as \cite{Ma11}:
\[
C_{GME} (\psi) = \min \lbrace \mathcal{C}_{1}, \mathcal{C}_{2}, \mathcal{C}_{3} \rbrace.
\]
With exception of $\xi$, these measures give 0 for bi-separable states. That is, $\xi$ does not discriminate bi-separability from genuine entanglement, so it is a measure of global entanglement. On the other hand, among these measures, $\mu$ averages the lowest results for the four states. In other words, $\mu$ defines a lower bound for measuring genuine entanglement.

To compare in more detail both the properties and the behavior of the measures $\xi$ and $\mu$, let us consider the type~3a state
\begin{equation}
\vert \psi_{\alpha} \rangle = \cos \alpha \vert 100 \rangle + \sin \alpha \vert 0 \rangle_1 \vert \Psi^+ \rangle_{23},  \quad 0 \leq \alpha \leq \tfrac{\pi}{2}.
\label{exa2}
\end{equation}
It is a matter of substitution to verify that $\vert \psi_{\alpha=0} \rangle =\vert 100 \rangle$ and $\vert \psi_{\alpha=\pi/2} \rangle =  \vert 0 \rangle_1 \vert \Psi^+ \rangle_{23}$. That is, at the ends of the $\alpha$-domain, the state $\vert \psi_{\alpha} \rangle$ becomes fully separable and bi-separable, respectively. Additionally, making $\alpha= \alpha_0= \arccos (\sfrac{1}{\sqrt 3})$, one gets $\vert \psi_{\alpha_0} \rangle = \vert W \rangle$.

%%%%%%%%%%%%%
\begin{figure}[htbp]
\centering
\subfloat[][]{\includegraphics[height=0.3\textwidth]{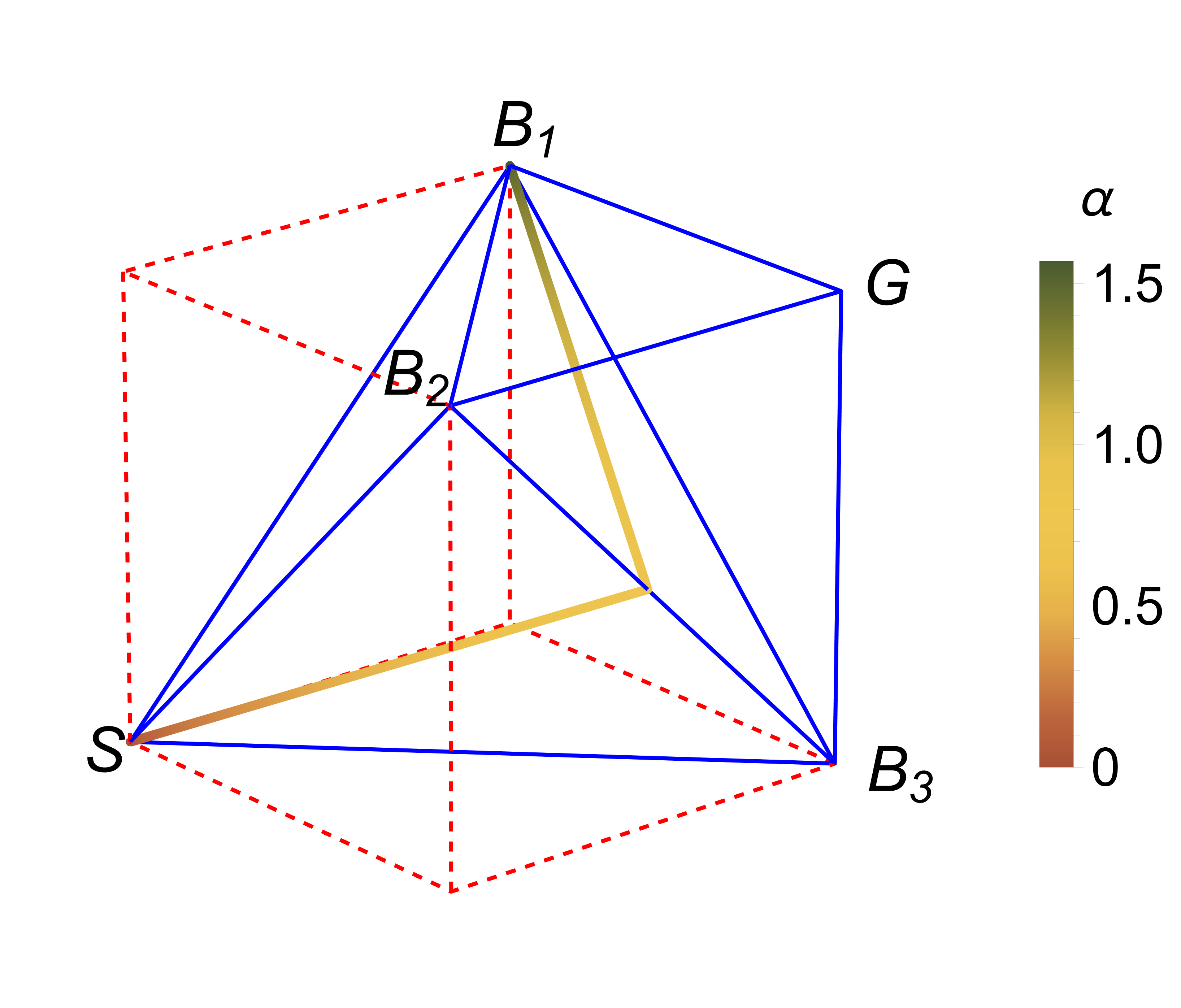}}
\hskip2cm
\subfloat[][]{\includegraphics[height=0.25\textwidth]{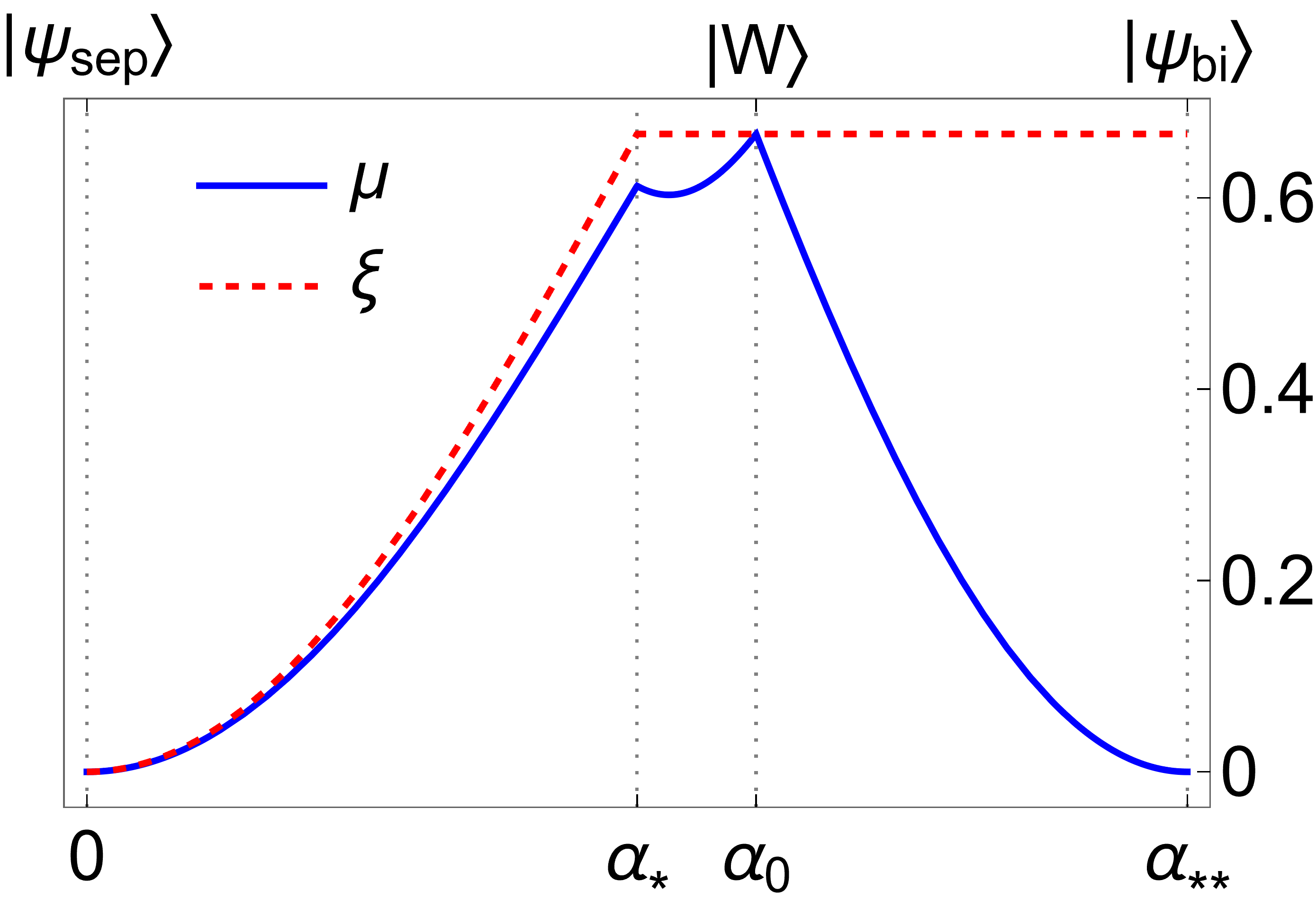}}

\caption{\footnotesize
(\textbf{a}) When the parameter $\alpha$ sweeps the domain $[0, \alpha_{**} ] \equiv [0, \pi/2]$, the state $\vert \psi_{\alpha} \rangle$ introduced in (\ref{exa2}) describes a path $\vec \lambda_{\psi_{\alpha}}$ on the lower tetrahedron of $\mathcal{P}$. The trajectory goes first from the vertex $\vec S$ to the midpoint of $\overline{B_{2}B_{3}}$ (at $\alpha= \alpha_* = \pi/4$), and then to the vertex $\vec B_1$ (at $\alpha = \alpha_{**}$) on the triangle $\triangle_{B_1B_2B_3}$. In this last journey, $\vec \lambda_{\psi_{\alpha}}$ passes through the geometric center of $\triangle_{B_1B_2B_3}$, at $\alpha= \alpha_0 = \arccos (\sfrac{1}{\sqrt 3})$, which houses state $\vert W \rangle$. (\textbf{b}) Measures $\xi$ and $\mu$ for state $\vert \psi_{\alpha} \rangle$, the former gives $\xi_0 = 2/3$ for $\alpha \in [\alpha_*, \alpha_{**}]$ whereas the latter cancels at the ends of the $\alpha$-domain. In contrast with $\xi$, the measure $\mu$ also distinguishes among the points living on $\triangle_{B_1B_2B_3}$. That is, points $\vec \lambda_{\psi_{\alpha}}$ closer to the center of $\triangle_{B_1B_2B_3}$ will correspond to more $\mu$-entangled states $\vert \psi_{\alpha} \rangle$.
}

\label{olaf}
\end{figure}
%%%%%%%%%%%%%

Parameterized by $\alpha$, the vector $\vec \lambda_{\psi_{\alpha}}$ describes a path in the entanglement--polytope $\mathcal P$ that follows a trajectory from $\vec S$ (at $\alpha=0$)  to $\vec B_1$ (at $\alpha= \pi/2$), see Figure~\ref{olaf}(a). Looping through values in the first half of the $\alpha$-domain, $\vec \lambda_{\psi_{\alpha}}$ draws a straight line on the triangle $\triangle_{SB_2B_3}$, from $\vec S$ to the midpoint of segment $\overline{B_2B_3}$ (at $\alpha = \pi/4$). From there, the vector continues in a straight line on triangle $\triangle_{B_1B_2B_3}$ to vertex $\vec B_1$. In this last journey, $\vec \lambda_{\psi_{\alpha}}$ passes through the geometric center of $\triangle_{B_1B_2B_3}$ (at $\alpha= \alpha_0$), which houses state $\vert W \rangle$.

Remember that mesure $\xi$ averages the constant value $\xi_0= 2/3$ for any point on the triangle $\triangle_{B_1B_2B_3}$, see Figure~\ref{xi}(b). Then, the values averaged by $\xi$ would indicate that the degree of entanglement of $\vert \psi_{\alpha} \rangle$ is conserved during the second part of the path described above, just as it is shown in Figure~\ref{olaf}(b). In turn, $\mu$ gives 0 for both separable and bi-separable states, so it cancels at the ends of the $\alpha$-domain. That is, with the values given by $\mu$ we find that $\vert \psi_{\alpha} \rangle$ increases its degree of entanglement during the first half of the journey, and decreases it during the second half to remain as at the beginning. According to $\mu$, the maximum degree of entanglement is reached at $\alpha= \alpha_0$, where $\vert \psi_{\alpha} \rangle$ coincides with $\vert W \rangle$. Indeed, $\mu$ averages two local maxima, the one at $\alpha=\alpha_0$, and another at $\alpha=\alpha_* = \pi/4$, see Figure~\ref{olaf}(b). Between $\alpha_*$ and $\alpha_0$ (the transit from $\overline{B_2B_3}$ to the geometric center of $\triangle_{B_1B_2B_3}$) there is still an increase of $\mu$ that connects these maxima. This last result shows that, in contrast with $\xi$, the measure $\mu$ also distinguishes among the points living on $\triangle_{B_1B_2B_3}$. The points $\vec \lambda_{\psi_{\alpha}}$ closer to the center of $\triangle_{B_1B_2B_3}$ will correspond to more $\mu$-entangled states $\vert \psi_{\alpha} \rangle$.

Let us verify the universality of measure $\mu$ to quantify genuine multipartite entanglement. Consider the state
\begin{equation} 
\vert \psi_{\theta} \rangle =  N(\theta) \left[  
\cos^2 \left( \tfrac{\theta}{2} \right) \vert 111\rangle + \tfrac{1}{\sqrt 3} \sin^2 \left( \tfrac{\theta}{2} \right) \vert W \rangle 
\right], \quad 
\theta \in [0, \pi],
\label{raro}
\end{equation}
where 
\[
N(\theta) = \left[
\cos^4 \left( \tfrac{\theta}{2} \right) + \tfrac13 \sin^4 \left( \tfrac{\theta}{2} \right)
\right]^{-1/2}
\]
stands for the normalization constant. It is immediate to verify the identities $\vert \psi_{\theta =0} \rangle = \vert 111 \rangle$, $\vert \psi_{\theta = \pi} \rangle = \vert W \rangle$ and $\vert \psi_{\theta = \theta_0} \rangle = \vert \varphi \rangle$, with $\theta_0 = 2 \pi/3$. The latter state is given in Eq.~(\ref{luGHZ}), which has been shown to be local unitary equivalent to the GHZ--state. The measures $\mu$ and $F_{123}$ are respectively given by
\begin{equation}
 \mu (\psi_{\theta}) = 1-\frac23 \bigg \vert \frac{3+5\cos \theta + \cos 2\theta}{3+2\cos \theta+\cos 2\theta}\bigg \vert,
 \label{muexample1}
\end{equation}
and
\begin{equation}
 F_{123}(\psi_{\theta})= \frac{8 (16 \cos \theta+ 5 \cos 2\theta +15)}{9(2\cos \theta +\cos 2\theta +3)} \sin^4 \left( \frac{\theta}{2} \right).
 \label{FEexample1}
\end{equation}
In turn, the GME--concurrence reads $\mathcal{C}_{\operatorname{GME}} (\psi_{\theta}) = \sqrt{F_{123}(\psi_{\theta})}$.

\begin{figure}[htbp]
\centering
\includegraphics[width=0.3\textwidth]{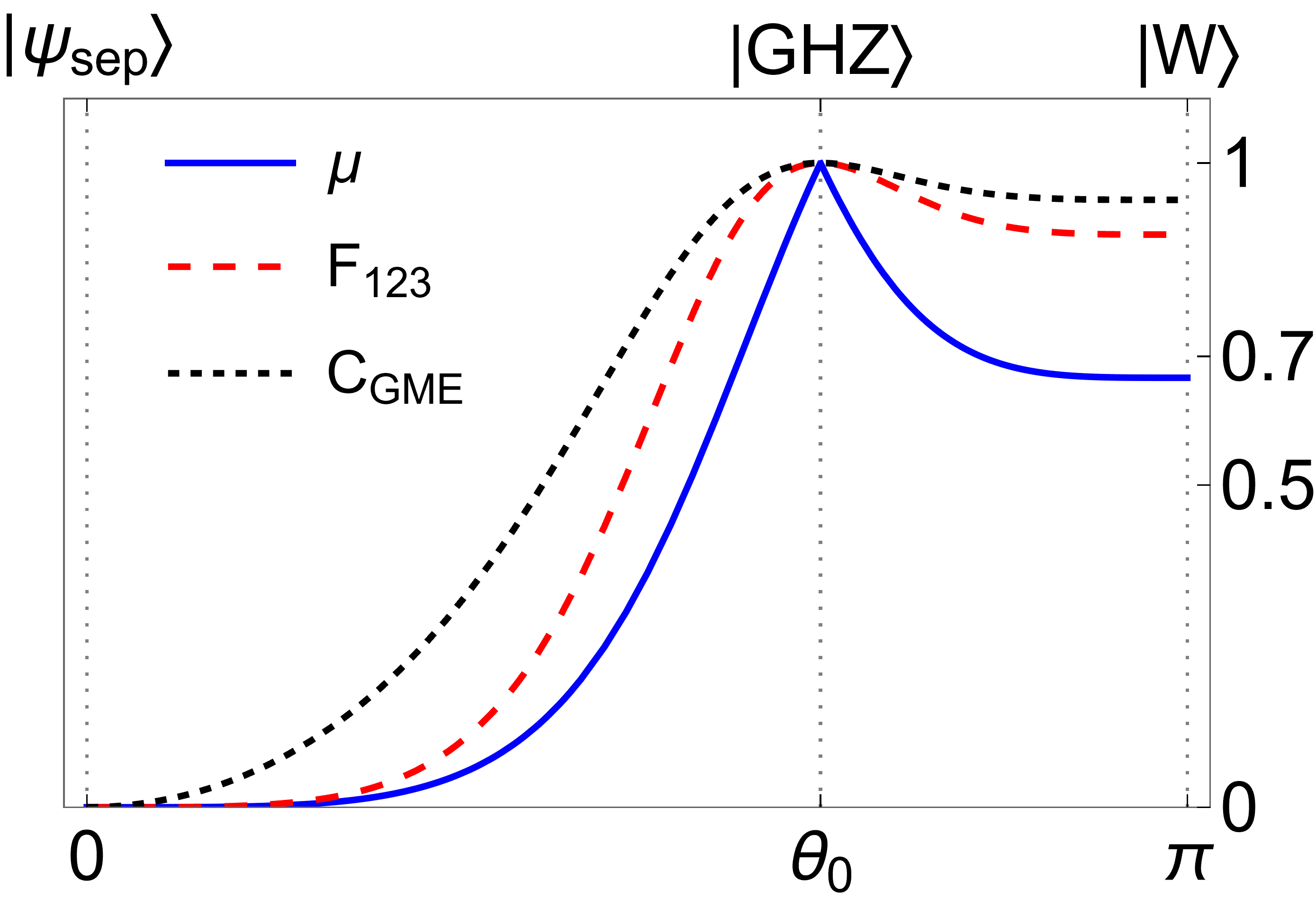}

\caption{\footnotesize
Entanglement measures $\mu(\psi)$, $F_{123} (\psi)$ and $\mathcal{C}_{\operatorname{GME}} (\psi)$ for the state $\vert \psi_{\theta} \rangle$ given in Eq.~(\ref{raro}). At $\theta =0$ the state becomes $\vert 111 \rangle$, so it is fully separable. However, it is maximally entangled at $\theta = \theta_0 = 2\pi/3$, since it is local unitary equivalent to the GHZ--state. At $\theta = \pi$ the state is equal to $\vert W \rangle$. Consistently, the measures range between 0 and 1, reaching their maximum ($=1$) at $\theta_0 $. Measure $\mu$ defines a lower bound, as it produces lower values than the results of the other two measures, except at $\theta= \theta_0$.
}

\label{solo}
\end{figure}

The above functions are plotted in Figure~\ref{solo}. They range between 0 and 1, with $\mu$ defining the lower bound for any $\theta \neq \theta_0$. At $\theta_0$, the three measures give the result 1.

%%%%%%%%%%%%%%
%%%%%%%%%%%%%%

Note that $\mu$ defines a lower bound on the degree of entanglement that could be measured in the state $\vert \psi_{\theta} \rangle$, as it produces lower values than the results of the other two measures.

%---------------------------------------> Section
\section{The inverse problem}
\label{inverse}

One of the advantages of working with the geometric representation of quantum states is that the classification of entanglement becomes visual and simple. If a given three-qubit state $\vert \psi \rangle \in \mathcal H$ exhibits a certain amount of entanglement, this is linked to a very concrete point of the entanglement--polytope $\mathcal P$ throughout the vector $\vec \lambda_{\psi}$. 

The question in the opposite direction offers even more interesting challenges and a much broader perspective on entanglement. 

Given a point $\vec \lambda \in \mathcal P$, what three-qubit state (or class of states) can be mapped from $\mathcal H$ to precisely $\vec \lambda$? 

This dilemma is an example of the inverse problem of quantum mechanics \cite{Fer97}, where one seeks to manipulate systems in order to force them to behave in a particular way (some applications can be found in \cite{Emm00,Ros03,Mie04,Cru07,Enr18,Enr19}). In the present case, it is about choosing a point of the polytope in a region that characterizes very specific entanglement properties, and searching for the quantum state that satisfies such a profile. 

Some of the possibilities of the inverse problem for states of types~1 to 4a have been explored in \cite{Han04} (although the authors of such work do not use the terminology of the inverse problem nor exploit all the potential of the method). However, as a matter of fact, the solutions to the inverse problem for three-qubit systems are far from being exhausted. For example, to the best of our knowledge, the inverse problem for the facets $B_{1}B_{2}G$, $B_{1}B_{3}G$ and $B_{2}B_{3}G$, remains unsolved.

To contribute to this topic, let us solve the problem of determining the states that can be associated with a given point $\vec \lambda$ on the triangle $\triangle_{B_{2}B_{3}G} \in \mathcal{P}$. 

The states that we are looking for belong to type~4c-1, so they are written in the form of Eq.~(\ref{state_4c1}). Since $\vec \lambda_{\psi_{4c-1}}$ must be provided in advance, we assume that we know every one of the coefficients of the convex combination (\ref{lambda_4c1}). Then, we solve the parameter system (\ref{kappas4c1}) to fix the coefficients of $\vert \psi_{4c-1} \rangle$ in (\ref{state_4c1}). The acceptable solutions are as follows
\begin{equation}
\begin{array}{l}
b_{2}^{2} = \frac{1}{4} \left( 1 + \kappa_{1}^{2} - \kappa_{2}^{2} - \sqrt{ \left( 1 - \kappa_{1}^{2} + \kappa_{2}^{2} \right)^{2} - 4 \kappa_{2}^{2} } \right),\\[3ex]
b_{3}^{2} = \frac{1}{4} \left( 1 - \kappa_{1}^{2} + \kappa_{2}^{2} - \sqrt{ \left( 1 - \kappa_{1}^{2} + \kappa_{2}^{2} \right)^{2} - 4 \kappa_{2}^{2} } \right),\\[3ex]
b_{4}^{2} = \frac{1}{2} \left( 1 - \sqrt{ \left( 1 - \kappa_{1}^{2} + \kappa_{2}^{2} \right)^{2} - 4 \kappa_{2}^{2} } \right).
\end{array}
\label{aes}
\end{equation}
Thus, coefficients (\ref{aes}) are determined by the parametrization $\kappa_1$, $\kappa_2$ and $1-\kappa_1 - \kappa_2$, which also defines  the facet $B_{2}B_{3}G$. 

The inverse problem for facets $B_{1}B_{3}G$ and $B_{1}B_{2}G$ finds a similar solution, this time using the states (\ref{state_ABG}).

In a more general picture, any point $\vec \lambda \in \mathcal P$ can be written as a convex combination of the extremal points,
\begin{equation}\label{eq:convexcom}
\vec{\lambda} = \kappa_{0} \vec{S} + \kappa_{1} \vec{B}_{1} + \kappa_{2} \vec{B}_{2} + \kappa_{3} \vec{B}_{3} + \kappa_{4} \vec{G}, \quad \kappa_q \geq 0, \quad \sum_{q=0}^{4} \kappa_{q} =1.
\end{equation}
This expression is universal for three-qubit systems in the sense that any of the states reported in Table~\ref{table1} can be associated with $\vec \lambda$. 

For example, knowing that states of type~4d are mapped to points along the entire entanglement--polytope $\mathcal{P}$, let us consider $\vert \psi_{4d} \rangle$ as it is given in Eq.~(\ref{SII25}). Making $c_{r}^{2} = v_{r}$, the corresponding smallest eigenvalues are
\begin{equation}
\begin{array}{ll}
 \lambda_{1} = \frac{1}{2} \left[ 1 - \sqrt{1 - 4 (v_{2}+v_{3}) \left( 1-v_{2}-v_{3} \right)}  \right],\\[1em]
 \lambda_{2} = \frac{1}{2} \left[ 1 - \sqrt{1 - 4  (v_{1}+v_{3}) \left( 1-v_{1}-v_{3} \right)}  \right],\\[1em]
 \lambda_{3} = \frac{1}{2} \left[ 1 - \sqrt{1 - 4 (v_{0}+v_{3}) \left( 1-v_{0}-v_{3} \right)}  \right].
\end{array}
\label{vs}
\end{equation}
Since these eigenvalues define the vector $\vec \lambda$, from (\ref{eq:convexcom}) and (\ref{vs}) we see that the coefficients of $\vert \psi_{4d} \rangle$ can be expressed in terms of $\kappa_i$, $i=0,1,2,3$, as follows
\begin{equation}\label{viequations}
\begin{array}{cc}
v_{0}  = \frac{1}{2} \left[ 1 - 2 v_{3} \pm (\kappa_{0}+\kappa_{3}) \right],\quad v_{1} = \frac{1}{2} \left[ 1 - 2 v_{3} \pm (\kappa_{0}+\kappa_{2}) \right],\\[1em]
v_{2} = \frac{1}{2} \left[ 1 - 2 v_{3} \pm (\kappa_{0}+\kappa_{1}) \right],
\quad 
v_3 + v_2 + v_1 + v_0 =1.
\end{array}
\end{equation}

Only four of the set of solutions above are admissible for our purposes. In particular, the case `$+$' yields
\begin{equation}
c_{0}^2 = \tfrac{1}{2} \left( \kappa_{3} + \tfrac{\kappa_4}{2} \right), \quad 
c_{1}^2 = \tfrac{1}{2} \left( \kappa_{2} + \tfrac{\kappa_{4}}{2} \right), \quad 
c_{2}^2 = \tfrac{1}{2} \left( \kappa_{1} + \tfrac{\kappa_{4} }{2} \right), \quad 
c_{3}^2= \tfrac{1}{2} \left( 1+\kappa_{0}- \tfrac{\kappa_{4} }{2} \right).
\label{SII34}
\end{equation}

The set (\ref{SII34}) constitutes the solution to the inverse problem of the entanglement--polytope $\mathcal P$ that we are dealing with. By providing purely geometric information, through the parameters $\kappa_i$, the state $\vert \psi_{4d} \rangle$ is completely determined, with very specific entanglement properties that can be defined on demand. 

To verify the universality of the above solution, first consider the vertex $\vec S$. That is, $\kappa_{0} = 1$. Then  $c_{3} = 1$ and $c_{0} = c_{1} = c_{2} = 0$. In this case we arrive at the fully separable state $\vert 111 \rangle$. Another immediate example arises if $\kappa_{4} = 1$ (the vertex $\vec G$), then $c_{0}=c_{1}=c_{2}=c_{3}= \sfrac12$, and
\begin{equation}
\vert \varphi \rangle = \frac12 \left( \sqrt{3} \vert W \rangle + \vert 111 \rangle \right).
\label{luGHZ}
\end{equation}
This state is local unitary equivalent to the GHZ state. Indeed, 
\[
\vert \varphi \rangle = \left( \sigma_x H \otimes \sigma_x H \otimes \sigma_x H \right) \vert GHZ \rangle,
\]
with $H$ the Hadamard operator. 

More interesting configurations are obtained when two or more $\kappa$-parameters are different from zero. In the extreme case, where $\kappa_i \neq 0$ for all $i=0,1,2,3,4$, one can pay attention to the barycenter of the simplex $\triangle_{B_1B_2B_3}$ (where state $\vert W \rangle$ is located). The inverse problem solution leads to the state
\begin{equation}
\label{otherW}
\vert \chi \rangle = \tfrac{1}{\sqrt 2}\left( \vert W \rangle + \vert 111 \rangle \right).
\end{equation}

It is remarkable that $\vert \chi \rangle$ is not local unitary equivalent to $\vert W \rangle$, although these two states are mapped into the same point of the entanglement--polytope $\mathcal P$. The difference is notable considering that the three-tangle of $\vert W \rangle$ is equal to zero \cite{Cof00}, while the three-tangle of $\vert \chi \rangle$ is equal to $4/(3\sqrt 3)$.

The above case shows the generality that the inverse problem introduces in the determination of quantum states. In the conventional (direct) problem, every state $\vert \psi \rangle \in \mathcal H$ is mapped to one and only one point of $\mathcal P$. However, the fact that two or more elements of $\mathcal H$ can be mapped to the same point of $\mathcal P$ usually goes unnoticed because, in the direct problem, attention is paid to a specific state. The inverse problem considers all possibilities in $\mathcal H$ that can be associated with $\vec \lambda \in \mathcal P$ in a single move. The latter means that the solution to the inverse problem usually associates a family of states, rather than a single state, with such a point.

As we have seen, the solution (\ref{SII34}) provides universality to the state $\vert \psi_{4d} \rangle$ in Eq.~(\ref{SII25}). Once the coefficients $c_k$ of $\vert \psi_{4d} \rangle$ are parametrized with purely geometric information, obtained from the entanglement--polytope $\mathcal P$, the entanglement properties of such state become manipulable. Furthermore, the strength of set (\ref{SII34}) lies in the fact that any other selection of solutions (\ref{viequations}) is local unitary equivalent to (\ref{SII34}). For example, taking the roots `$+, -, -$', from (\ref{viequations}) we have
\[
 c_0^2 =\tfrac{1}{2} \left( 1+\kappa_{0}- \tfrac{\kappa_{4} }{2} \right), \quad 
 c_1^2 =\tfrac{1}{2} \left( \kappa_{1} + \tfrac{\kappa_{4} }{2} \right), \quad 
 c_2^2 =\tfrac{1}{2} \left( \kappa_{2} +  \tfrac{\kappa_{4} }{2} \right), \quad 
 c_3^2=\tfrac{1}{2} \left( \kappa_{3} + \tfrac{\kappa_{4} }{2} \right).
\]
The state $\vert \widetilde\psi_{4d} \rangle$ that results from these coefficients is local unitary equivalent to $\vert \psi_{4d} \rangle$ through the transformation $U_1=\sigma_{x} \otimes \sigma_{x} \otimes \mathds{1}$. Similarly, the roots `$-,+,-$' and `$ -,-,+$' provide states of type~4d that can be transformed into the form $\vert \psi_{4d} \rangle$ by using the unitary operators $U_2=\sigma_{x} \otimes \mathds{1} \otimes \sigma_{x}$ and $U_3=\mathds{1} \otimes \sigma_{x} \otimes \sigma_{x}$, respectively.

We have chosen type~4d states to exemplify, in a more or less general way, the applicability and power of the inverse problem. Therefore, we must emphasize that this method is applicable to any point $\vec \lambda_{\psi} \in \mathcal{P}$, in connection with the states included in Table~\ref{table1}. 

\begin{figure}[htbp]
\centering
\subfloat[][]{\includegraphics[height=0.3\textwidth]{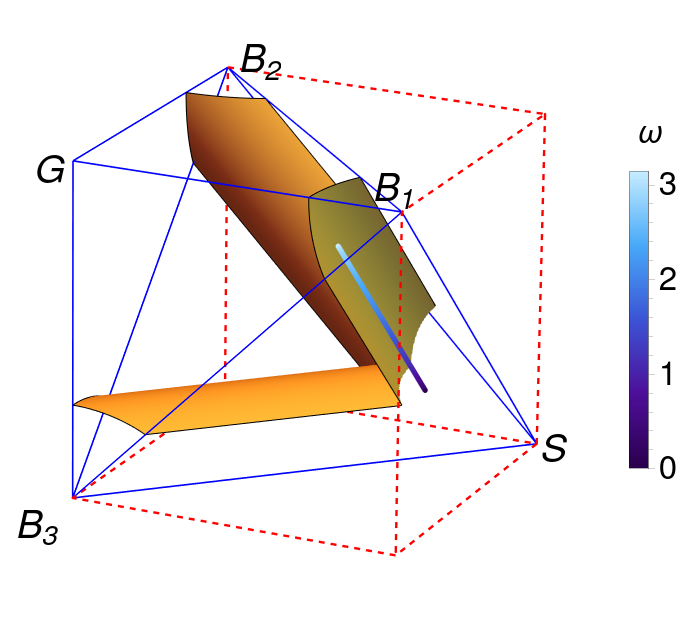}}
\hskip2cm
\subfloat[][]{\includegraphics[height=0.25\textwidth]{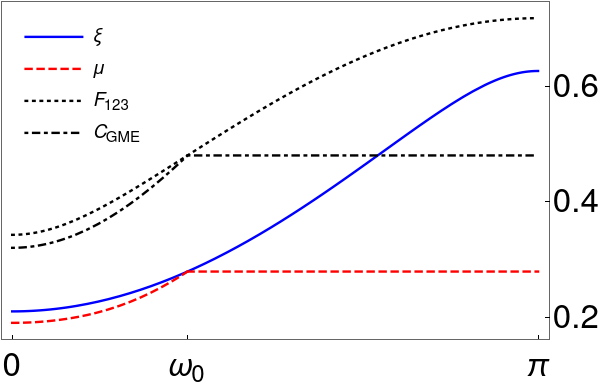}}

\caption{\footnotesize 
A particular solution to the inverse problem of finding a set of points in $\mathcal P$ that follows a path with entanglement measure $\mu= \mu_0 = \operatorname{const}$, see Figure~\ref{mutopo}(b), is obtained after making $b_{\ell} = 1/\sqrt{5}$ in (\ref{standard}). The pure state $\vert \psi \rangle$ is parameterized by the phase $0 \leq \omega \leq \pi$ and is projected onto the path $\vec \lambda_{\psi(\omega)}$ in $\mathcal P$ . (\textbf{a}) $\vec \lambda_{\psi(\omega)}$ starts and finishes at $\left( \sfrac{\mu_{0}}{2}, \varepsilon, \varepsilon \right)$ and $\left( \sfrac{\mu_{0}}{2}, \sfrac25, \sfrac25 \right)$, respectively. Here $\mu_{0} = 1 - \sfrac{\sqrt{13}}{5}$ and $\varepsilon = \tfrac{1}{2} \left( 1 - \sfrac{\sqrt{17}}{5} \right)$. In the interval $[0, \omega_0]$, with $\omega_0 = \pi/3$, the path overflows the $\mu_0$-region.
(\textbf{b}) Measures $\mu$, $\xi$, $F_{123}$ and $\mathcal{C}_{GME}$ as functions of $\omega$. Only $\mu$ and $\mathcal{C}_{GME}$ maintain constant in the interval $[\omega_0, \pi]$, with values $\mu = 1 - \sfrac{\sqrt{13}}{5}$ and $\mathcal{C}_{GME} = \sfrac{12}{25}$.
}
	
\label{Fomega}
\end{figure}

We can also pose the inverse problem in another context. Suppose you are interested in determining a collection of points that describe a given path in $\mathcal P$. Say, the path follows one of the surfaces determined by $\mu = \mu_0 =\operatorname{const}$. It can be shown that a particular solution is obtained after making $b_{\ell} = 1/\sqrt{5}$, $\ell =0,1,2,3,4$, in (\ref{standard}). Looping through all the domain $[0, \pi]$ of $\omega$, we obtain the path shown in Figure~\ref{Fomega}(a). In a given interval of $\omega$, the path overflows the $\mu_0$-region towards vertex $S$. That is, entanglement is constant for $\omega \in [\omega_0, \pi]$ and decreases as $\omega$ goes from $\omega_0$ to zero. Our assertion is verified in Figure~\ref{Fomega}(b), where we show the measures $\mu$, $\xi$, $F_{123}$ and $\mathcal{C}_{GME}$ for the $\omega$-dependent solution we have found to the above inverse problem. 

%%%%%%%%%%%%%
%%%%%%%%%%%%%

In general, the versatility of the solutions of the inverse method could extend to the practical aspects of entanglement. For example, the creation and measurement of three-qubit entanglement associated with states of  type~ 4c-2 have been reported in \cite{Agu15}. Since these results are parametrized by data obtained from the optical bench in the laboratory, information like this could be translated to the entanglement--polytope $\mathcal P$ to give it measurable properties.

%---------------------------------------> Section
\section{Conclusions}
\label{conclu}

We have studied the degree of entanglement between the different parts of a tripartite qubit system from a purely geometric perspective. After constructing the convex polytope $\mathcal P \subset \mathbb{R}^3$ formed by the points $\vec \lambda = (\lambda_1, \lambda_2, \lambda_3)$, where $\lambda_k$ is the smallest eigenvalue of the reduced matrix associated with the $k$th qubit \cite{Han04,Wal13,Lun20,Hig03}, we have considered the map $\Lambda: \mathcal{H} \rightarrow \mathcal{P}$ to identify some relationships between the tripartite quantum states $\vert \psi \rangle \in \mathcal{H}$ and the points of the projective space $\cal P$. 

In agreement with the conditions reported in \cite{Han04}, we have shown that the classification of entangled states introduced in \cite{Aci00,Aci01} results in the identification of concrete subsets of $\mathcal{P}$ under the mapping $\Lambda$. The emphasis in the present study has been on the states mapped into the facets of the polytope $\mathcal P$.

Considering the geometric properties of the polytope $\mathcal P$, we have introduced two different entanglement measures, denoted $\xi$ and $\mu$. They are respectively associated with the projection and rejection of the point $\vec \lambda$ on the biseparable segments of $\mathcal P$; the former quantifies global entanglement while the latter measures genuine entanglement. When compared with some previously reported measures of entanglement, it is found that $\xi$ and $\mu$ establish a lower bound for the type of entanglement to which they refer (excluding the GHZ--state for which both measurements return the value 1, as would be expected).

The above results can be extended in several directions. For example, one may consider the projection of $\vec \lambda$ onto different subsets of $\mathcal P$ in such a way that different entanglement information is provided. In general, the definition of $\xi$ and $\mu$ can be directly extended to the multi-qubit case, since the points of the corresponding entanglement--polytope (the dimension of the space that contains it does not matter) can always be projected onto the appropriate subsets of biseparable points.

The advantages of working with the geometric representation of quantum states, as we have done here, become more evident when considering the possibility of controlling and manipulating entanglement. 

As we have shown, by solving the inverse problem we can force the system to behave in a particular way. In particular, given a point of the polytope in a region that characterizes very specific entanglement properties, the quantum state that satisfies such a profile is sought. As a matter of fact, the solution to the inverse problem associates a family of states in $\mathcal H$, rather than a single state, with such a point. The latter provides information of the space of states $\mathcal H$  that cannot be obtained by solving the conventional (direct) problem, where a given state in $\mathcal H$ is
mapped to one and only one point of $\mathcal P$.

Setting the value of $\xi$ or $\mu$ identifies regions of the entanglement--polytope $\mathcal P$ whose points represent states with exactly the same degree of entanglement. These regions allow us to presuppose various evolutions of the tripartite system, associated with trajectories on some hypersurface of $\mathcal P$, which are characterized by leaving the degree of initial entanglement invariant. The inverse problem allows us to determine the type of operations that must be applied on the states of the system to induce said trajectories in $\mathcal P$. Even better, it opens the possibility of inducing an increase in the degree of entanglement by identifying operations that correspond to transitions between the different hypersurfaces of $\mathcal P$ that are characterized by the entanglement measures $\xi$ and $\mu$. Work in this direction is in progress and will be published elsewhere. 

%---------------------------------------> Section
\appendix
\section{Three-qubit states in standard form}
\label{ApA}

\renewcommand{\thesection}{A-\arabic{section}}
% redefine the command that creates the equation no.
\setcounter{section}{0}  % reset counter 

\renewcommand{\theequation}{A-\arabic{equation}}
% redefine the command that creates the equation no.
\setcounter{equation}{0}  % reset counter 

\renewcommand{\thetable}{A-\arabic{table}}
% redefine the command that creates the equation no.
\setcounter{table}{0}  % reset counter 

Let $\mathcal B_{\mathcal H} = \{ \vert i_1, i_2, i_3 \rangle, i_k =0,1, k=1,2,3\}$ be the orthonormal basis of the Hilbert space  $\mathcal H = \mathcal H_2 \otimes \mathcal H_2 \otimes \mathcal H_2$. The standard form allows any three-qubit pure state $\vert \psi \rangle \subset \mathcal H$ to be described using the minimal number of product states $\vert i_1, i_2, i_3 \rangle$. According with \cite{Aci00,Aci01}, this requirement is satisfied by considering the orthonormal subset of basis vectors
\[
 \mathcal B_{\mathcal H} \supset
 \mathcal S_{\mathcal P} = \left\{ \vert 000 \rangle, \vert 100 \rangle, \vert 101 \rangle, \vert 110 \rangle, \vert 111 \rangle
\right\}.
\]
We have written $\vert \psi \rangle$ as the linear superposition (\ref{standard}). The phase $0 \leq \omega \leq \pi$, which is linked to the basis vector $\vert 100 \rangle$, plays a relevant role if $b_{\ell} \neq 0$, $\ell = 1,2,3,4$. Indeed, after diagonalizing, the reduced one-qubit density matrices $\rho_k = \operatorname{diag} (\lambda_k, 1- \lambda_k)$ can be parameterized by the smallest eigenvalue $\lambda_k = (1 - \sqrt{1-4 \det \rho_k})/2$, $k=1,2,3$, where
\begin{equation}
\begin{array}{c}
\det \rho_{1} = b_{0}^{2} \left( 1 - b_{0}^{2} - b_{1}^{2}  \right),
\\[1ex]
\det \rho_{2} = b_{3}^{2} \left( b_{0}^{2} + b_{2}^{2} \right) + b_{4}^{2} \left( b_{0}^{2} + b_{1}^{2} \right) - 2b_{1}b_{2}b_{3}b_{4} \cos \omega,
\\[1ex]
\det \rho_{3} = b_{2}^{2} \left( b_{0}^{2} + b_{3}^{2} \right) + b_{4}^{2} \left( b_{0}^{2} + b_{1}^{2} \right) - 2b_{1}b_{2}b_{3}b_{4} \cos \omega.
\end{array}
\label{omega}
\end{equation}

The classification discussed in Section~\ref{qualitative} obeys the separability properties of $\vert \psi \rangle$, as it is written in Eq.~(\ref{standard}), in connection with the different subsets of $\mathcal S_{\mathcal P}$. 

Let $\mathcal S_{\mathcal P}^{(m)} \subseteq \mathcal S_{\mathcal P}$ be a subset composed of $m$ elements $\vert i_1, i_2, i_3 \rangle$, its size is determined by the binomial coefficient $\binom{5}{m}$. This gives rise to five different subsets: $\mathcal S_{\mathcal P}^{(1)}$ and $\mathcal S_{\mathcal P}^{(4)}$ with 5 elements, $\mathcal S_{\mathcal P}^{(2)}$ and $\mathcal S_{\mathcal P}^{(3)}$ with 10 elements, and $\mathcal S_{\mathcal P}^{(5)}$ with only one element.

Given $\mathcal S_{\mathcal P}^{(m)}$, not all its elements have been included in Table~\ref{table1} of Section~\ref{qualitative} since we seek a classification of the states in terms of their amount of entanglement, which is preserved under local-unitary transformations. In this sense, if two or more elements of the same $\mathcal S_{\mathcal P}^{(m)}$ are local-unitary equivalent, then we are left with only one of them. The same criterion applies if there exist local-unitary equivalent states between different subsets. 

The set $\mathcal S_{\mathcal P}^{(1)}$ is composed of the fully separable states $\vert i_1, i_2, i_3 \rangle$, which are local-unitary equivalent with each other, so we can take $\vert 000 \rangle$ as the generic case.  

The classification of fully separable states is not exhausted in the set $\mathcal S_{\mathcal P}^{(1)}$ since some elements of $\mathcal S_{\mathcal P}^{(2)}$ also belong to that class. In fact, five elements of $\mathcal S_{\mathcal P}^{(2)}$ generate separable pure states, see Table~\ref{tableA1}. They have also been represented by the generic basis vector $\vert 000 \rangle$ in Table~\ref{table1}, since there is a local-unitary transformation in $U(1) \times SU(2) \times SU(2) \times SU(2)$ that maps every one of them into the form $\vert 000 \rangle$ \cite{Aci00,Aci01}. The remaining five elements give rise to states of type~2, but the combination $\{ \vert 100 \rangle, \vert 111 \rangle \}$ belongs to the orbit of $\{ \vert 101 \rangle, \vert 110 \rangle \}$, which is the representative state of type~2a-1. That is, $\mathds{1} \otimes \mathds{1} \otimes \sigma_{x}  \left( b_{1} \vert 100 \rangle + b_{4} \vert 111 \rangle \right) = b_{1} \vert 101 \rangle + b_{4} \vert 110 \rangle$. Therefore, Table~\ref{table1} includes only four distinguishable elements of $\mathcal S_{\mathcal P}^{(2)}$, those representing generic states of type~2.

%%%%%%%%%%%%%%%%%%%%%%%%%%%
\begin{table}[h!]
	\centering
	\scalebox{1}{
		\begin{tabular}{lll @{\vrule height 12pt depth 2pt width 0pt}}
			\hline
			Type & Basis product states & Subset $\subseteq \mathcal P$ \\ \hline
			1 &	$\lbrace \vert 000 \rangle, \vert 100 \rangle \rbrace$ & $\vec S$ \rule[-9pt]{0pt}{12pt}\\
			1 & $\lbrace \vert 100 \rangle, \vert 101 \rangle \rbrace$ & $\vec S$ \rule[-9pt]{0pt}{12pt}\\
			1 &	$\lbrace \vert 100 \rangle, \vert 110 \rangle \rbrace$ & $\vec S$ \rule[-9pt]{0pt}{12pt}\\
			1 & $\lbrace \vert 101 \rangle, \vert 111 \rangle \rbrace$ & $\vec S$ \rule[-9pt]{0pt}{12pt}\\
			1 & $\lbrace \vert 110 \rangle, \vert 111 \rangle  \rbrace$ & $\vec S$ \rule[-9pt]{0pt}{12pt}\\
			2a-1 & $\lbrace \vert 100 \rangle, \vert 111 \rangle \rbrace$  & $\overline{SB_{1}}$  \rule[-9pt]{0pt}{12pt}\\ 
			\hline
			2a-1 & $\lbrace \vert 101 \rangle, \vert 110 \rangle\rbrace$ & $\overline{SB_{1}}$ \rule[-9pt]{0pt}{12pt}\\
			2a-2 & $\lbrace \vert 000 \rangle, \vert 101 \rangle \rbrace$ & $\overline{SB_{2}}$  \rule[-9pt]{0pt}{12pt}\\
			2a-3 &	$\lbrace \vert 000 \rangle, \vert 110 \rangle \rbrace$ & $\overline{SB_{3}}$  \rule[-9pt]{0pt}{12pt}\\
			2b & $ \lbrace \vert 000 \rangle, \vert 111 \rangle \rbrace$ & $\overline{SG}$ \rule[-9pt]{0pt}{12pt}\\
			\hline
		\end{tabular}
}
	
\caption{\footnotesize Classification of the pure states $\vert \psi \rangle$ generated by the different elements of $\mathcal S_{\mathcal P}^{(2)}$. From top to bottom, the first five rows correspond to fully separable (type~1) states, generically represented by $\vert 000 \rangle$ in Table~\ref{table1}. The sixth row corresponds to states that are local-unitary equivalent to the seventh row states. The last four rows refer to states of type~2 that are completely distinguishable from each other, providing non-redundant information in Table~\ref{table1}.
}
\label{tableA1}
\end{table}
%%%%%%%%%%%%%%%%%%%%%%%%%%%

In Table~\ref{tableA2} we see that six elements of $\mathcal S_{\mathcal P}^{(3)}$ provide states that can be transformed into any of the generic type~2 states described above. The remaining four elements yield the generic type~3 states that are included in Table~\ref{table1}.

%%%%%%%%%%%%%%%%%%%%%%%%%%%
\begin{table}[h!]
	\centering
	\scalebox{1}{
		\begin{tabular}{lll @{\vrule height 12pt depth 2pt width 0pt}}
			\hline
			Type & Basis product states & Subset $\subseteq \mathcal P$ \\ \hline
2a-1 & $\lbrace \vert 100 \rangle, \vert 101 \rangle, \vert 110 \rangle \rbrace$  & $\overline{SB_{1}}$ \rule[-9pt]{0pt}{12pt}\\
			2a-1 & $\lbrace \vert 100 \rangle, \vert 101 \rangle, \vert 111 \rangle \rbrace$  & $\overline{SB_{1}}$ \rule[-9pt]{0pt}{12pt}\\
			2a-1 & $\lbrace \vert 100 \rangle, \vert 110 \rangle, \vert 111 \rangle \rbrace$  & $\overline{SB_{1}}$ \rule[-9pt]{0pt}{12pt}\\
			2a-1 & $\lbrace \vert 101 \rangle, \vert 110 \rangle, \vert 111 \rangle \rbrace$ & $\overline{SB_{1}}$ \rule[-9pt]{0pt}{12pt}\\
			2a-2 & $\lbrace \vert 000 \rangle, \vert 100 \rangle, \vert 101 \rangle \rbrace$  & $\overline{SB_{2}}$  \rule[-9pt]{0pt}{12pt}\\
			2a-3 & $\lbrace \vert 000 \rangle, \vert 100 \rangle, \vert 110 \rangle \rbrace$  & $\overline{SB_{3}}$  \rule[-9pt]{0pt}{12pt}\\
			\hline
			3a & $\lbrace \vert 000 \rangle, \vert 101 \rangle, \vert 110 \rangle \rbrace$  & Facets of $SB_{1}B_{2}B_{3}$ \rule[-9pt]{0pt}{12pt}\\
			3b-1 & $\lbrace \vert 000 \rangle, \vert 100 \rangle, \vert 111 \rangle \rbrace$ & $SB_{1}G$\\
			3b-2 & $\lbrace \vert 000 \rangle, \vert 101 \rangle, \vert 111 \rangle \rbrace$  & $SB_{2}G$\\ 
			3b-3 & $\lbrace \vert 000 \rangle, \vert 110 \rangle, \vert 111 \rangle \rbrace$  &  $SB_{3}G$ \rule[-9pt]{0pt}{12pt}\\
			\hline
		\end{tabular}
}
	
\caption{\footnotesize Classification of the pure states $\vert \psi \rangle$ generated by the different elements of $\mathcal S_{\mathcal P}^{(3)}$. From top to bottom, the first six rows correspond to states that can be transformed into the indicated generic states of type~2. The last four rows correspond to the generic type~3 states included in Table~\ref{table1}.
}
\label{tableA2}
\end{table}
%%%%%%%%%%%%%%%%%%%%%%%%%%%

The set $\mathcal S_{\mathcal P}^{(4)}$ deserves particular attention. On the one hand, 
it includes the element $\{ \vert 1, i_2, i_3 \rangle \}$, which requires $b_{\ell} \neq 0$, $\ell =1,2,3,4$. From (\ref{omega}) we see that the smallest eigenvalues $\lambda_{1,2}$ of the states generated by this element are necessarily parameterized by $\omega$. Therefore, is in this case that $\omega$ becomes relevant. Nevertheless, $\{ \vert 1, i_2, i_3 \rangle \}$ is not of type~4 but of type 2a-1, see Table~\ref{tableA3}. On the other hand, the remaining elements of $\mathcal S_{\mathcal P}^{(4)}$ correspond to the generic type~4a, 4b and 4c states included in Table~\ref{table1}.

%%%%%%%%%%%%%%%%%%%%%%%%%%%
\begin{table}[h!]
	\centering
	\scalebox{1}{
		\begin{tabular}{lll @{\vrule height 12pt depth 2pt width 0pt}}
			\hline
			Type & Basis product states & Subset $\subseteq \mathcal P$ \\ \hline
2a-1 & $\lbrace \vert 100 \rangle, \vert 101 \rangle, \vert 110 \rangle, \vert 111 \rangle \rbrace$ & $\overline{SB_{1}}$ \rule[-9pt]{0pt}{12pt}\\
\hline
			4a & $\lbrace \vert 000 \rangle, \vert 100 \rangle, \vert 101 \rangle, \vert 110 \rangle \rbrace$  & $SB_{1}B_{2}B_{3}$ \rule[-9pt]{0pt}{12pt}\\
			4b-2 & $\lbrace \vert 000 \rangle, \vert 100 \rangle, \vert 101 \rangle, \vert 111 \rangle \rbrace$  &  $\subset SB_{1}B_{2}G$  \rule[-9pt]{0pt}{12pt}\\	
			4b-1 & $\lbrace \vert 000 \rangle, \vert 100 \rangle, \vert 110 \rangle, \vert 111 \rangle \rbrace$  & $\subset SB_{1}B_{3}G$\\
			4c & $\lbrace \vert 000 \rangle, \vert 101 \rangle, \vert 110 \rangle, \vert 111 \rangle \rbrace$  & $SB_{1}B_{2}B_{3} \cup SB_{2}B_{3}G$ \rule[-9pt]{0pt}{12pt}\\
\hline
		\end{tabular}
}
	
\caption{\footnotesize Classification of the pure states $\vert \psi \rangle$ generated by the different elements of $\mathcal S_{\mathcal P}^{(4)}$. From top to bottom, the first row corresponds to states that can be transformed into the generic type~2a-1 state. The last four rows correspond to the generic type~4a, 4b and 4c states included in Table~\ref{table1}.
}
\label{tableA3}
\end{table}
%%%%%%%%%%%%%%%%%%%%%%%%%%%

The set $\mathcal S_{\mathcal P}^{(5)}$ requires $b_{\ell} \neq 0$ for all $\ell$. This is associated with a subset of $\mathcal P$ which is disjoint with the subsets linked to the generic states of types~1, 2, 3 and 4 described above (otherwise, $b_{\ell} =0$ for some $\ell$).

Finally, note that the type~4d state included in Table~\ref{table1} is not generated by $\mathcal S_{\mathcal P} \subset \mathcal B_{\mathcal H}$ but by another subset of $\mathcal B_{\mathcal H}$, which is composed of four elements and has only two elements in common with $\mathcal S_{\mathcal P}$. Following \cite{Car00}, the related pure states $\vert \psi_{4d} \rangle$ have been written as the linear superposition (\ref{SII25}).

%---------------------------------------> Section
\section*{Author Contributions}

Conceptualization, M.E. and O.R.-O.; methodology, formal analysis, investigation and original draft preparation, M.E., S.L.-H. and O.R.-O.; review and edditing O.R.-O.; project administration and funding acquisition, O.R.-O. All authors have read and agreed to the published version of the manuscript.

%---------------------------------------> Section
\section*{Funding}

This research was funded by Consejo Nacional de Humanidades, Ciencia y Tecnolog\'ia (CONAHCyT, Mexico), grant number A1-S-24569, and by Instituto Politécnico Nacional (IPN, Mexico), project SIP20242277.

%---------------------------------------> Section
\section*{Acknowledgments}

S. Luna-Hern\'andez acknowledges the support from CONAHCyT through the scholarship 592045.

%---------------------------------------> Section
\section*{Conflicts of Interest}

The authors declare no conflict of interest.

%---------------------------------------> Bibliography

\end{document}